\documentclass[12pt]{article}
\pdfoutput=1
\usepackage{graphicx}
\usepackage{amsmath, amssymb}
\usepackage{slashed}
\usepackage{xcolor}
\usepackage{array,arydshln}
\usepackage{cite}
\usepackage[pdftex,
  colorlinks=true,
citecolor=green!60!blue]{hyperref}

\def\baselinestretch{1.3}

\newcommand{\rem}[1]{{\bf #1}}
\newcommand{\sla}[1]{\not \! \! #1}

\begin{document}

\begin{titlepage}

\def\thefootnote{\fnsymbol{footnote}}

\begin{center}

\hfill KANAZAWA-19-02 \\
\hfill June, 2019

\vspace{0.5cm}
{\Large\bf Leptogenesis \\ after superconformal subcritical hybrid inflation}

\vspace{1cm}
{\large Yoshihiro Gunji}, 
{\large Koji Ishiwata}

\vspace{1cm}

{\it Institute for Theoretical Physics, Kanazawa University, Kanazawa
  920-1192, Japan}







\vspace{1cm}

\abstract{We consider an extended version of superconformal
  subcritical hybrid inflation model by introducing three right-handed
  neutrinos that have the Majorana mass terms. In the model one of the
  right-handed sneutrinos plays a role of the inflaton field, and it
  decays to reheat the universe after inflation. The vacuum expectation
  value for the waterfall field gives an unconventional pattern of the
  light neutrino mass matrix, and the neutrino Yukawa couplings that
  determine the reheating temperature are not constrained by the
  neutrino oscillation data. Consequently thermal leptogenesis or
  sneutrino leptogenesis is realized. }

\end{center}
\end{titlepage}

\renewcommand{\theequation}{\thesection.\arabic{equation}}
\renewcommand{\thepage}{\arabic{page}}
\setcounter{page}{1}
\renewcommand{\thefootnote}{\#\arabic{footnote}}
\setcounter{footnote}{0}

\section{Introduction}
\label{sec:intro}
\setcounter{equation}{0} 

Inflation paradigm is strongly supported by the observations of the
cosmic microwave background radiation (CMB). Slow-roll scalar field in
the early universe is a promising candidate for inflation, and many
types of inflation models have been proposed so far.  In a theoretical
point of view, it would be tempting to ask what the underlying physics
or symmetry of the inflaton field is.  Supersymmetry (SUSY) might be
one of the answers.  It protects the flatness of the inflaton
direction, which is suitable for inflation.

Recently supersymmetric D-term hybrid inflation has been revisited in
various point of view. Under shift symmetric K\"ahler
potential~\cite{Kawasaki:2000yn}, subcritical hybrid inflation was
found, where inflation continues for subcritical point value of the
inflaton field~\cite{Buchmuller:2014rfa,Buchmuller:2014dda}. On the
other hand, it was shown in
Refs.\,\cite{Buchmuller:2012ex,Buchmuller:2013zfa} that Starobinsky
model~\cite{Starobinsky:1980te} emerges in the framework of
superconformal
supergravity~\cite{Einhorn:2009bh,Kallosh:2010ug,Ferrara:2010yw,Ferrara:2010in}.
It turned out in the following study that this framework has another
new regime of inflation.  It was shown that a general class of
superconformal $\alpha$-attractor
model~\cite{Kallosh:2013yoa,Kallosh:2013xya} appears in the
subcritical regime of inflation, which we call superconformal
subcritical hybrid inflation~\cite{Ishiwata:2018dxg}.  In addition,
the energy scale of inflation should coincide with the grand
unification scale to be consistent with the Planck observation data,
which is the feature found in the subcritical hybrid
inflation~\cite{Buchmuller:2014rfa,Buchmuller:2014dda}. Namely, the
superconformal subcritical hybrid inflation has both features of the
superconformal $\alpha$-attractor models and the subcritical hybrid
inflation. The shift symmetry and superconformality are crucial for
them.

In this paper we will study the thermal history after the end of
superconformal subcritical hybrid inflation. (See
Refs.~\cite{Bryant:2016tzg,Bryant:2016sjj} that study the
phenomenology of Pati-Salam version of subcritical hybrid
inflation. Recently Ref.~\cite{Domcke:2017rzu} comprehensively studies
the D-term hybrid inflation, including reheating, leptogenesis, and
the SUSY breaking mechanism.) For the purpose, we introduce three
right-handed neutrinos that interact with the minimal supersymmetric
standard model (MSSM) sector. In fermionic sector, the mass matrix for
the light neutrinos is given by the seesaw mechanism~\cite{seesaw},
but it has an unconventional structure.  In bosonic sector, on the
other hand, it will be shown that one of sneutrinos can play a role of
the inflaton field. In addition, baryon asymmetry that is sufficient
amount to explain the observed value is generated via
leptogenesis~\cite{Fukugita:1986hr}.

This paper is organized as follows. In the next section the model that
we consider is described. Then Sec.\,\ref{sec:inf} shows the
conditions required for the superconformal subcritical hybrid
inflation in this model.  Mass matrices of the heavy and light
(s)neutrinos, including parametrization of the neutrino Yukawa
couplings, are given in Sec.\,\ref{sec:neu_mass}, then we discuss the
reheating and leptogenesis after inflation in
Sec.\,\ref{sec:post_inf}. Sec.\,\ref{sec:conclusion} is dedicated to
conclusion.

\section{The model}
\label{sec:model}
\setcounter{equation}{0}

We consider a model described by the superpotential
\begin{align}
  W=W_{\rm MSSM}+W_{\rm neu}\,,
\end{align}
where $W_{\rm MSSM}$ is the superpotential of the MSSM sector and 
\begin{align}
  W_{\rm neu}=
  \frac{1}{2}M_{ij}N^c_iN^c_j+y_{\nu\,ij}N^c_i L_jH_u
  +\lambda_i N_i^c S_+ S_-\,.
  \label{eq:Wneu}
\end{align}
Here $N_i^c$, $L_i=(\nu_{Li},l_{Li})^T$, and $H_u=(H_u^+,H_u^0)^T$ are
the chiral superfields of the right-handed neutrinos, left-handed
leptons, and up-type Higgs, respectively, and
$L_jH_u=\nu_{Lj}H_u^0-l_{Lj}H_u^+$. $S_{\pm}$ are the local U(1)
fields with charge $\pm q$ $(q>0)$, one of which plays a role of the
waterfall field.\footnote{We write the superpartners with tilde for
  the MSSM fields and right-handed neutrinos. For $S_\pm$, the same
  symbols are used for scalar fields while fermionic parts are
  expressed with tilde. In the current and the next sections, we
  adopt the unit in which the reduced Planck mass $M_{pl}\simeq
  2.4\times 10^{18}\,{\rm GeV}$ is taken to be unity unless otherwise
  mentioned. } Indices are summed over $i,j=1$\,--\,3. (We will use
the same contraction in the following discussion unless otherwise
mentioned.) $M_{ij}$ terms are explicit superconformal breaking terms
that are added by phenomenological purpose. While we do not argue
their origin, $M_{ij} \ll 1$ are expected. The K\"{a}hler potential,
on the other hand, is given by
\begin{align}
  {\cal K}=-3\log \Omega^{-2}\,,
\end{align}
where 
\begin{align}
  \Omega^{-2}=1-\frac{1}{3}
  \left(\sum_{\rm MSSM} |I_{\rm MSSM}|^2+\sum_i |N_i^c|^2+|S_+|^2+|S_-|^2\right)
  -
  \sum_i\frac{\chi_i}{6}\left(N_i^{c\,2}+\bar{N}_i^{c\,2}\right)\,.
\end{align}
Here $I_{\rm MSSM}$ are chiral superfields in the MSSM sector. The
last term is superconformal breaking term that is considered in
Refs.~\cite{Buchmuller:2012ex,Ferrara:2010in}. With the superpotential
and K\"ahler potential, the scalar potential is given by
\begin{align}
  V_{\rm tot}=V_F+V_D,
\end{align}
where $V_F$ and $V_D$ are F- and D-terms, respectively, and given
by~\cite{Buchmuller:2012ex}
\begin{align}
  &V_F=\Omega^4\left[
    \delta^{\alpha\bar{\alpha}}W_\alpha W_{\bar{\alpha}}
    +\frac{1}{\Delta}
    |\delta^{\alpha\bar{\alpha}}W_\alpha \Phi_{\bar{\alpha}}-3W|^2
    \right]\,,\\
  &V_D=\frac{1}{2}g^2\left[q\Omega^2(|S_+|^2-|S_-|^2)-\xi\right]^2\,.
\end{align}
Here $\Phi \equiv -3\Omega^{-2}$ and $\Delta \equiv
\Phi-\delta^{\alpha\bar{\alpha}}\Phi_\alpha\Phi_{\bar{\alpha}}$ have
been additionally introduced.  Subscript in $W$ and $\Phi$ stands for
the field derivative, {\it e.g.}, $W_\alpha\equiv \partial W/\partial
z^\alpha$ where $z^\alpha$ is a chiral superfield. In the D-term, we
have introduced the Fayet-Iliopoulos (FI) term $\xi\,(>0)$ associated
with the U(1).\footnote{The origin of the FI term in canonical
  superconformal supergravity model~\cite{Ferrara:2010in} is discussed
  in Ref.\,\cite{Buchmuller:2012ex}. See also
  Ref.\,\cite{Domcke:2017rzu} for recent development. } Due to the FI
term, $S_+$ has a vacuum expectation value (VEV) at the global
minimum, which is obtained as $\langle S_+
\rangle=\sqrt{\xi/q(1+\tilde{\xi})}$ with $\tilde{\xi}\equiv \xi/3q$.

As in Ref.\,\cite{Ishiwata:2018dxg}, we take $\chi_i \le 0$ without
loss of generality. In the present model, we impose the following
condition:
\begin{align}
  \chi_3\simeq -1,~~ \chi_{1},\,\chi_2\simeq 0\,.
\end{align}
This distinguishes $N^c_3$ from $N^c_{1}$ and $N^c_{2}$. $\phi \equiv
\sqrt{2}{\rm Re}\,\tilde{N}^c_3 $ has an approximate shift symmetry
that is explicitly broken by $\lambda_3\,(\ll 1)$. Then, $\phi$ is
expected to be the inflaton as studied in
Ref.\,\cite{Ishiwata:2018dxg}.  In the inflation model, $ s \equiv
\sqrt{2}|S_+|$ plays a role of the waterfall field. $N^c_1$ and
$N^c_2$, on the other hand, have no such symmetry.  Instead there is a
freedom to choose any basis for $N^c_1$ and $N^c_2$ with a
redefinition of $M_{ij}$ and $\lambda_i$ due to $\chi_{1,2}\simeq 0$.

For simple notation, hereafter we omit subscripts of $\chi_3$ and
$\lambda_3$ and introduce $m_\phi$, which will be identified as the
inflaton mass in Sec.\,\ref{sec:subcriticalregime} (with another
assumption in Sec.\,\ref{sec:mass_matrix}),
\begin{align}
  \chi \equiv \chi_3, &~~ \lambda \equiv \lambda_3\,, \\
  m_\phi \equiv&\, \lambda \langle S_+ \rangle \,.
 \label{eq:m_phi}
\end{align}
It is sometimes convenient to use $\delta \chi$ ($0<\delta \chi<1$)
defined by $\delta \chi/(1+\chi) = -qg^2\xi/3\lambda^2$. $0<\delta
\chi<1$ guarantees $\Omega(\phi,s)^2>0$ and $\phi_{c,0}^2>0$, which
will be defined in Eqs.\,\eqref{eq:Omega(phi,s)} and
\eqref{eq:phi_c0}, respectively.  Then inflation that is consistent
with the observations of the CMB is realized in the parameter
space~\cite{Ishiwata:2018dxg},
\begin{align}
  \lambda\simeq
  (0.5\, {\rm \mathchar`-}{\rm \mathchar`-}\,1)\times 10^{-3}\,, &~~
  \xi^{1/2}\simeq
  (3\, {\rm \mathchar`-}{\rm \mathchar`-}\,1)\times 10^{16}\,{\rm GeV}\,,
  \nonumber \\
  m_{\phi} \simeq (1\,{\rm \mathchar`-}{\rm \mathchar`-}\,2)\times&
  10^{13}\,{\rm GeV}\,,
  \label{eq:parameter_space}
\end{align}
for $q=g=1$, $\delta \chi=0.9$ and the number of $e$-folds
$N_e=55$\,--\,60, which we take in the later numerical
study.\footnote{We will estimate the number of $e$-folds in
  Sec.\,\ref{sec:reheating} to confirm this. } In the parameter space,
{\it e.g.}, $\chi\simeq -1.16$ for $\lambda=10^{-3}$,
$\xi^{1/2}=10^{16}\,{\rm GeV}$, and $\delta \chi=0.9$. The mass terms
in the superpotential, however, have a possibility to alter the
inflationary path.  In the next section, we will derive the conditions
in order not to affect the inflationary dynamics.

\section{Inflation}
\label{sec:inf}
\setcounter{equation}{0} 

We define several variables that are used in the following
analysis. During inflation, the other fields except for the inflaton
and waterfall field are irrelevant. Thus it is convenient to define
following potentials,
\begin{align}
 & V(\phi,s)\equiv V_{\rm tot}|_{\sqrt{2}{\rm Re}\,\tilde{N}_3^c
    =\phi,\,\sqrt{2}|S_+|=s,\,{\rm the\,others}=0}\,,
  \\
 & \Omega(\phi,s)\equiv \Omega|_{\sqrt{2}{\rm Re}\,\tilde{N}_3^c
    =\phi,\,\sqrt{2}|S_+|=s,\,{\rm the\,others}=0}\,.
  \label{eq:Omega(phi,s)}
\end{align}
Then the critical point value $\phi_c$ is defined as a field value
below which the waterfall field becomes tachyonic. It receives ${\cal
  O}(M_{ij}^2)$ corrections as
\begin{align}
  \phi_c^2=\phi_{c,0}^2\left[1-\frac{2\Delta M^2(\phi_{c,0})}{3\lambda^2}\right]
  +{\cal O}(M_{i3}^4)\,,
  \label{eq:phi_c}
\end{align}
where  
\begin{align}
  \phi_{c,0}^2&=\frac{6qg^2\xi}{3\lambda^2+(1+\chi)qg^2\xi}\,,
  \label{eq:phi_c0}\\
  \Delta M^2(\phi) &= \sum_{i=1}^3|M_{i3}|^2
    -\frac{(1-2\chi)^2}{24}\frac{\phi^2}{1+\phi^2\chi(1+\chi)/6}|M_{33}|^2\,.
\end{align}
For example, $\phi_{c,0}\simeq 18$ or in terms of
canonically-normalized field $\hat{\phi}_{c,0}\simeq 11$ defined in
Eq.\,\eqref{eq:phi2phiHat} for $\lambda=10^{-3}$,
$\xi^{1/2}=10^{16}\,{\rm GeV}$ and $\delta \chi=0.9$. This
perturbative expansion is valid when\footnote{We have checked that
  ${\cal O}(M^4_{i3})$ term is irrelevant when
  Eq.\,\eqref{eq:cond_for_phic_expnd2} is satisfied, thus we ignore it
  in the following discussion.}
\begin{align}
  |M_{13}|^2+|M_{23}|^2 &\ll 3\lambda^2/2
  \sim (10^{15}\,{\rm GeV})^2\,,
  \label{eq:cond_for_phic_expnd1} \\
  |M_{33}|^2 &\ll 2\lambda^4/qg^2\xi\sim (10^{14}\,{\rm GeV})^2\,.
  \label{eq:cond_for_phic_expnd2}
\end{align}
It will be checked in this section that the above conditions are
satisfied in this inflation model.

Finally it is useful to define
\begin{align}
   \Psi\equiv \frac{\Omega(\phi,0)\phi}{\Omega(\phi_{c,0},0)\phi_{c,0}}
=\frac{\Omega(\phi,0)\phi}{\sqrt{2qg^2\xi/\lambda^2}}\,,
\end{align}
when potential is expressed in terms of canonically-normalized
inflaton field.

\subsection{Pre-critical regime}

Let us begin with the regime where the inflaton is approaching down to
the critical point value. Since the waterfall field is stabilized at
the origin in this regime, the relevant Lagrangian is given as
\begin{align}
  {\cal L}_{\rm pre}=
  \frac{f(\phi,0)}{2}(\partial_\mu \phi)^2-V_{\rm pre}(\phi)\,,
\end{align}
where\footnote{It is noted that the term proportional to $\Delta M^2$
  in Eq.\,\eqref{eq:Vpre} is equivalent to
  $\Omega^4(\phi,0)\phi^2\Delta M^2(\phi)/2$.}
\begin{align}
  f(\phi,s)&=\Omega^2(\phi,s)
  \left[1+\Omega^2(\phi,s)\frac{(1+\chi)^2}{6}\phi^2\right]\,, \\
  V_{\rm pre}(\phi)&=
  \frac{1}{2}g^2\xi^2\left[
    1+2\Psi^2 \frac{\Omega^2(\phi,0)\Delta M^2(\phi)}{\lambda^2\xi/q}
    \right]\,.
  \label{eq:Vpre}
\end{align}
It is seen that $\Delta M^2$ term gives a gradient to the inflaton
field, which should not invade the slow-roll conditions. To see the
impact of $\Delta M^2$ term, it is instructive to change dynamical
variable $\phi$ to canonically-normalized field $\hat{\phi}$.  Since
$\chi\simeq -1$, it is a good approximation that $f(\phi,0)\simeq
\Omega^2(\phi,0)$. Then $d\phi/d\hat{\phi}=f(\phi)^{-1/2}\simeq
\Omega^{-1}(\phi,0)$ can be solved easily to obtain,
\begin{align}
  &\phi \simeq \beta^{-1/2}\sinh\beta^{1/2}\hat{\phi}\,,
  \label{eq:phi2phiHat} \\
  &\Psi \simeq \delta \chi^{-1/2}\tanh\beta^{1/2}\hat{\phi}\,, \\
  &\Omega^{-1}(\phi,0) \simeq \cosh\beta^{1/2}\hat{\phi}\,,
  \label{eq:Omg^-1inphiHat}
\end{align}
where $\beta=-(1+\chi)/6=\lambda^2\delta \chi/2qg^2\xi$. Then
the potential in terms of $\hat{\phi}$  is given as
\begin{align}
  V_{\rm pre} &\simeq \frac{1}{2}g^2\xi^2 
  \biggr[1+
    \frac{\tanh^2\beta^{1/2}\hat{\phi}}{\cosh^2\beta^{1/2}\hat{\phi} }
    \frac{2q}{\lambda^2\xi\delta\chi} 
  \biggr\{
    \sum_{i=1}^3|M_{i3}|^2- \frac{3}{8\beta}|M_{33}|^2
    \tanh^{2}\beta^{1/2}\hat{\phi}
    \biggl\}\biggl]\,.    
\end{align}
On the other hand, it was shown in Ref.\,\cite{Buchmuller:2012ex} that
there is one-loop corrections to the tree-level potential. In terms of
the canonically-normalized field, it is given by
\begin{align}
  V_{1l}\simeq \frac{1}{2}g^2\xi^2
  \times\frac{q^2g^2}{8\pi^2}
  \log\left[\delta\chi^{-1}\tanh^2\beta^{1/2}\hat{\phi}\right]\,.
\end{align}
Therefore, in order not to affect the inflationary trajectory, it is
sufficient that the terms proportional to $|M_{i3}|^2$ are subdominant
compared to the one-loop potential. In the parameter space given in
Eq.\,\eqref{eq:parameter_space}, $\sqrt{\beta}\hat{\phi_c}\simeq
\mathrm{arcsinh} \sqrt{\beta}\phi_c \sim {\cal O}(1)$. Then, the
conditions are given as
\begin{align}
  |M_{13}|^2+|M_{23}|^2  &\lesssim
  \frac{qg^2\delta \chi\lambda^2\xi}{16\pi^2}
  \lesssim (1\times 10^{12}\,{\rm GeV})^2\,,
  \label{eq:constraint_on_M13_M23}\\
  |M_{33}|^2 &\lesssim
  \frac{\delta \chi^2\lambda^4}{12\pi^2}
  \lesssim (2\times 10^{11}\,{\rm GeV})^2\,.
  \label{eq:constraint_on_M33}
\end{align}
It is easy to check that the slow-roll conditions are satisfied under
the constraints.  Since the constraints are more stringent than
Eqs.\,\eqref{eq:cond_for_phic_expnd1} and
\eqref{eq:cond_for_phic_expnd2}, it has been confirmed that the
perturbative expansion to obtain Eq.\,\eqref{eq:phi_c} is valid.

\subsection{Subcritical regime}
\label{sec:subcriticalregime}

In the previous subsection, we have seen that the slow-roll conditions
are satisfied before reaching to the critical point value. After the
inflaton field becomes subcritical point value, the tachyonic growth
of the waterfall field occurs. It is expected that the inflation
continues in the subcritical regime when $M_{i3}\to 0$. In this
subsection, we will derive the conditions under which the inflaton and
waterfall field dynamics are not affected with non-zero $M_{i3}$.

As seen in the previous section, the perturbative expression for
$\phi_c$ is valid under the conditions given in
Eqs.\,\eqref{eq:constraint_on_M13_M23} and
\eqref{eq:constraint_on_M33}. Then, the tachyonic growth of the
waterfall field is not affected by the additional gradient in the
inflaton direction due to $|M_{i3}|^2$ terms since $\phi_c\simeq
\phi_{c,0}$. Consequently, the dynamics of the waterfall field around
the critical point is the same as one discussed in
Ref.\,\cite{Ishiwata:2018dxg}. The tachyonic growth is qualitatively
the same as the subcritical hybrid
inflation~\cite{Buchmuller:2014rfa,Buchmuller:2014dda}. Namely, due to
the tachyonic growth, the waterfall field relaxes to the local minimum
value $s_{\rm min}$ just after a few Hubble-unit time. (See, for
example, Figure 1. of Ref.\,\cite{Buchmuller:2014rfa}.) In the present
model, $s_{\rm min}$ is found to be
\begin{align}
  s_{\rm min}^2
  &=\frac{2\xi}{q(1+\tilde{\xi})}
  \frac{\Omega^{-2}(\phi,0)}{1+\tilde{\xi}\Psi^2/(1+\tilde{\xi})}
  \left[1-\Psi^2
    \left\{1+\frac{2\Omega^2(\phi,0)\Delta M^2(\phi)}{3\lambda^2} \right\}
    \right]\,.
  \label{eq:smin}
\end{align}
Then the potential in the subcritical regime of the inflaton field is
effectively given by $V(\phi,s_{\rm min})$ and the dynamics reduces to
single field inflation that is described by the Lagrangian,
\begin{align}
  {\cal L}=
  \frac{f(\phi,s_{\rm min})}{2}(\partial_\mu \phi)^2- V_{\rm inf}(\phi)\,,
\end{align}
where
\begin{align}
  V_{\rm inf}(\phi)&=
  g^2\xi^2\frac{(1+\tilde{\xi})\Psi^2}{1+2\tilde{\xi}\Psi^2}
  \biggl[1-\frac{\Psi^2}{2(1+\tilde{\xi})} 
    +
    \frac{\{1+\tilde{\xi}\Psi^2/(1+\tilde{\xi})\}^2}
         {(1+2\tilde{\xi}\Psi^2)/(1+\tilde{\xi})}
    \frac{\Omega^2(\phi,0)\Delta M^2(\phi)}{\lambda^2\xi/q}
    \biggr]\nonumber \\ &+{\cal O}(M_{i3}^4) \,.
\end{align}
Recall that $\tilde{\xi}\,,\xi\ll 1$ in the parameters in
Eq.\,\eqref{eq:parameter_space}. Then it is clear that the additional
term proportional to $\Delta M^2$ is the same as in
Eq.\,\eqref{eq:Vpre}. Note that $s_{\rm min}$ is negligible in
$\Omega(\phi,s_{\rm min})$. Then, the canonically-normalized inflaton
field is given by
Eqs.\,\eqref{eq:phi2phiHat}--\eqref{eq:Omg^-1inphiHat}.  Consequently,
$V_{\rm inf}$ is given by
\begin{align}
  V_{\rm inf} &\simeq g^2\xi^2 \delta \chi^{-1} \tanh^2\beta^{1/2} \hat{\phi}
  \biggr[1-\frac{\delta \chi^{-1}}{2} \tanh^2\beta^{1/2}
    \hat{\phi} \nonumber \\
    &+\frac{\cosh^{-2}\beta^{1/2}\hat{\phi}}{\lambda^2\xi/q}
    \biggr\{
    \sum_{i=1}^3|M_{i3}|^2-\frac{3}{8\beta}|M_{33}|^2
    \tanh^{2}\beta^{1/2}\hat{\phi}
    \biggl\}\biggl]\,.
\end{align}
Therefore, if Eqs.\,\eqref{eq:constraint_on_M13_M23} and
\eqref{eq:constraint_on_M33} are satisfied, then the dynamics in the
subcritical regime reduces to one in Ref.\,\cite{Ishiwata:2018dxg}.
In addition, it is clear that $m_\phi$ defined in
Eq.\,\eqref{eq:m_phi} is indeed the inflaton mass since $V_{\rm
  inf}\simeq \frac{1}{2}m_\phi^2 \phi^2$ around $\phi\,(\simeq
\hat{\phi})\sim 0$.

To summarize the present and previous subsections, the
inflaton-waterfall field dynamics is unchanged when
\begin{align}
  |M_{i3}|^2=|M_{3i}|^2 \lesssim (2\times 10^{11}\,{\rm GeV})^2\,,
  \label{eq:constraint_on_Mi3}
\end{align}
for $i=1$--3 are satisfied.

\subsection{Stability of inflationary trajectory}

It was pointed out in Ref.\,\cite{Nakayama:2016gvg} that $\tilde{L}_i
H_u$ may become tachyonic in sneutrino inflation. In order to find out
the stability condition, let us derive the mass matrix in
$\tilde{L}_i$ and $H_u$ basis.  From $V_{\rm tot}$, it is obtained by
\begin{align}
  {\cal L}_{\rm mass}^{\tilde{L}\mathchar`- H_u}=&
  -\frac{\Omega(\phi,s)^4 \phi^2}{2} |y_{\nu 3i} \tilde{L}_i|^2
  -\frac{\Omega(\phi,s)^4 \phi^2}{2} |y_{\nu 3i}|^2|H_u|^2
  \nonumber \\
  &+\left[\frac{M_{33}^* \Omega(\phi,s)^6\phi^3}{4\sqrt{2}}
    y_{\nu 3i} \tilde{L}_iH_u
+{\rm h.c.}
    \right]\,,
\end{align}
On the other hand, the kinetic terms of $\tilde{L}_i$ and $H_u$ are
given by
\begin{align}
  {\cal L}_{\rm kin}^{\tilde{L}\mathchar`- H_u}=
  \Omega(\phi,s)^2\left[|\partial_\mu \tilde{L}_i|^2+|\partial_\mu H_u|^2\right]\,.
\end{align}
Therefore, using canonically-normalized fields,
$\hat{\tilde{L}}_i\equiv \Omega(\phi,s)\tilde{L}_i$ and
$\hat{H}_u\equiv \Omega(\phi,s)H_u$, the mass terms are rewritten as
\begin{align}
  {\cal L}_{\rm mass}^{\tilde{L}\mathchar`- H_u}=&
  -\frac{\Omega(\phi,s)^2 \phi^2}{2} |y_{\nu 3i} \hat{\tilde{L}}_i|^2
  -\frac{\Omega(\phi,s)^2 \phi^2}{2} |y_{\nu 3i}|^2|\hat{H}_u|^2
  \nonumber \\
  &+\left[\frac{M_{33}^* \Omega(\phi,s)^4\phi^3}{4\sqrt{2}}
    y_{\nu 3i} \hat{\tilde{L}}_i\hat{H}_u
    +{\rm h.c.}
    \right]
  \nonumber \\
  =&
  -\frac{\Omega(\phi,s)^2 \phi^2}{2}
  (y_\nu y_\nu^\dagger)_{33}|\hat{\tilde{L}}_3^\prime|^2
  -\frac{\Omega(\phi,s)^2 \phi^2}{2}(y_\nu y_\nu^\dagger)_{33} |\hat{H}_u|^2
  \nonumber \\
  &+\left[\frac{M_{33}^* \Omega(\phi,s)^4\phi^3}{4\sqrt{2}}
    (y_\nu y_\nu^\dagger)_{33}^{1/2}
    \hat{\tilde{L}}_3^\prime\hat{H}_u
    +{\rm h.c.}
    \right]\,,
\end{align}
where we have defined $(y_\nu
y_\nu^\dagger)_{33}^{1/2}\hat{\tilde{L}}_3^\prime \equiv y_{\nu 3i}
\hat{\tilde{L}}_i$ in the second line following
Ref.\,\cite{Nakayama:2016gvg}. Then, the stability condition is given by
\begin{align}
  |M_{33}|
  <2\sqrt{2}(y_\nu y_\nu^\dagger)_{33}^{1/2}\frac{\Omega(\phi,s)^{-2}}{\phi}\,.
\end{align}
Using Eqs.\,\eqref{eq:phi2phiHat} and \eqref{eq:Omg^-1inphiHat}, it
turns out that
\begin{align}
  |M_{33}|&<4\sqrt{2}(y_\nu y_\nu^\dagger)_{33}^{1/2}\sqrt{\beta}
  \nonumber \\
  &\simeq 1.4 \times 10^{14}\,{\rm GeV}
  \left(\frac{(y_\nu y_\nu^\dagger)_{33}}{10^{-7}}
  \frac{\beta}{10^{-3}}\right)^{1/2}\,.
  \label{eq:stability}
\end{align}
This upper bound is weaker than \eqref{eq:constraint_on_Mi3} in most of the
parameter space, which will be seen later.

\section{Neutrino mass}
\label{sec:neu_mass}
\setcounter{equation}{0}

In this section, we derive the mass matrices for the heavy and light
neutrinos. Around the global minimum, $\Omega \simeq 1$ since $\xi\ll
1$.  Consequently, all the fields are canonical. Thus, the mass terms
are derived similarly in global SUSY model.

\subsection{Mass matrix}
\label{sec:mass_matrix}

The superpotential \eqref{eq:Wneu} gives the Majorana masses for the
light neutrinos. To see how the masses are generated, we write down
the mass terms for fermionic part of $N_i^c$, $\nu_{Li}$ and
$\tilde{S}_-$,
\begin{align}
  {\cal L}_{\nu}^{\rm mass}=
  -\frac{1}{2}(\bar{\psi}{\cal M}P_L\psi+{\rm h.c.})\,,
\end{align}
where
\begin{align}
  \psi&=(N^c_1,N^c_2,N^c_3,\tilde{S}_-,\nu_{L1},\nu_{L2},\nu_{L3})^T\,,
  \\
  {\cal M}&=\left(
\begin{array}{cc}
  \tilde{M} & \tilde{m}_\nu \\
  \tilde{m}_\nu^T & {\bf 0} 
\end{array} 
\right)\,.
\end{align}
$\tilde{M}$ and $\tilde{m}_\nu$ are $4\times 4$ and $4\times 3$
matrices, respectively, and given by
\begin{align}
  \tilde{M}&=
      \left(
\begin{array}{cccc}
&&&\lambda_1\langle S_+\rangle \\
  &\mbox{\smash{\Large $M$}}&&\lambda_2\langle S_+\rangle\\
&&& m_\phi\\ 
  \lambda_1\langle S_+\rangle &\lambda_2\langle S_+\rangle
  &m_\phi  & 0
\end{array}
\right)\,, \\
\tilde{m}_\nu&=
      \left(
\begin{array}{ccc}
&& \\
  &\mbox{\smash{\Large $m_\nu$}}&\\
&& \\ 
 0&0&0
\end{array}
\right)\,.
\end{align}
Here $m_{\nu\,ij}=y_{\nu\,ij}\langle H_u^0\rangle$ with $\langle
H_u^0\rangle$ being the VEV of the up-type neutral Higgs.  Then mass
matrix $M_{\nu}$ for the light neutrinos are obtained by the seesaw
mechanism~\cite{seesaw},
\begin{align}
 M_{\nu} =-\tilde{m}_\nu^T \tilde{M}^{-1}\tilde{m}_\nu\,.
\end{align}
An important consequence of the mass matrix is that the one of three
light neutrinos is massless. This is because the rank of $M_\nu$ is
two. Using this mass matrix, it is possible to constrain the
parameters by the observed neutrino masses.

In the later discussion we assume
\begin{align}
  \lambda_1\,,~\lambda_2\,\ll \lambda\,.
  \label{eq:lam_{1,2}=0}
\end{align}
Here, recall that there is a freedom to choose a basis for $N^c_1$ and
$N^c_2$. Then, $M_{12}$ can be rotated away. As a result, $M_{\nu}$ is
given in the following simple expression,
\begin{align}
  M_{\nu\, ij} = -\langle H^0_u\rangle^2
  \sum_{k=1}^2\frac{y_{\nu ki}y_{\nu kj}}{M_{kk}}
  +{\cal O}\left(\frac{\lambda_{1,2}}{\lambda}
  \frac{m_{\nu\, il}m_{\nu\, jm}}{M_{kk}}\right)\,.
  \label{eq:Mnu_lim}
\end{align}
Here in the second term of right-hand side, $k=1$ or $2$ and $i$, $j$,
$l$, $m$ can be 1--3, and we have ignored $M_{i3}$ based on the
discussion of the previous section. The leading order term, on the
other hand, is independent of $\lambda_{1,2}$, $M_{i3}$, $m_\phi$ and
$y_{\nu 3i}$ ($i=1$--3). Therefore, they are not constrained by the
neutrino oscillation data. This fact is important in the estimation of
the reheating temperature, which we will see later. To ensure the
non-zero $\lambda_{1,2}$ does not affect our later analysis at percent
level, we implicitly assume $\lambda_{1,2}<0.01\lambda$.

Before further discussing the light neutrino mass matrix, let us note
that the mass matrix $\tilde{M}$ corresponds to the mass matrix in the
superpotential around the global minimum,
\begin{align}
  W_{\rm neu}=\frac{1}{2} (N^c_1,N^c_2,N^c_3,S_-)\tilde{M}
  \left(
  \begin{array}{c}
    N^c_1 \\
    N^c_2 \\
    N^c_3 \\
    S_- 
    \end{array}
  \right)
  +\cdots \,.
\end{align}
Recall that we have the requirement \eqref{eq:constraint_on_Mi3} for
successful inflation. Therefore, $\tilde{M}$ should be almost
block-diagonal as
\begin{align}
  \label{eq:Mtilde}
  \tilde{M}&\simeq
      \left(
\begin{array}{ccc:c}
M_1&&&0 \\
  &M_2&&0\\
&& M_3& m_\phi\\
\hdashline
  0 &0&m_\phi & 0
\end{array}
\right)\,,
\end{align}
where $M_i>0$. Here we have left $M_3$ for later discussion. In the
following analysis we use Eq.\eqref{eq:Mtilde} for
$\tilde{M}$. $\tilde{M}$ is further diagonalized by a unitary matrix
$U_{\tilde{M}}$ as,
\begin{align}
   D_{\tilde{M}}\simeq
        {\rm diag}(M_1,M_2,
       m_\phi, m_\phi)\simeq
        U^T_{\tilde{M}} \tilde{M} U_{\tilde{M}}\,,
\end{align}
where
\begin{align}
    U_{\tilde{M}}=
  \left(
  \begin{array}{cc:c}
    1 &   &  \\
      & 1 &  \\
    \hdashline
        & &
        {\scriptsize
          \frac{1}{\sqrt{2}}\bigl(
          \begin{array}{cc}
            1 & i\\
            1 & -i
          \end{array}
          \bigr)}
  \end{array}
  \right)\,.
\end{align}
Here we have omitted ${\cal O}(M_3/m_\phi)$ corrections since they are
irrelevant in the later analysis.  Non-zero $\lambda_{1,2}$
corrections enter as ${\cal O}(\lambda_{1,2}/\lambda)$ and ${\cal
  O}((m_\phi/M_{1,2})\lambda_{1,2}/\lambda)$ for $m_\phi\gtrsim
M_{1,2}$ and $m_\phi \lesssim M_{1,2}$, respectively.

The corrections due to non-zero $\lambda_{1,2}$ are similar in the
scalar sector since the mass matrix in
$(\tilde{N}_1,\tilde{N}_2,\tilde{N}_3,S_-)$ basis is given by,
\begin{align}
 M_{\rm scalar}^2=\tilde{M}^\dagger \tilde{M}\,.
\end{align}
Therefore the assumption \eqref{eq:lam_{1,2}=0} assures that the
dynamics of the inflaton and the waterfall field discussed in the
previous section, as well as subsequent reheating and leptogenesis, is
not affected. If it is not satisfied, then the inflationary trajectory
would be altered so that we need further analysis to find the paramter
space that is consistent with the CMB observations. Accordingly, the
subsequent reheating and leptogenesis scenario change. We leave the
detailed analysis as an interesting future work.
  
\subsection{Parametrization of neutrino Yukawa couplings}

\begin{table}[t]
 \begin{center}
  \begin{tabular}{ccc}
   \hline \hline
   & $\Delta m_{21}^2\,[10^{-5}\,{\rm eV}^2]$ &
   $\Delta m_{3l}^2\,[10^{-3}\,{\rm eV}^2]$\\
   \hline
   NH &$7.39^{+0.21}_{-0.20}$ &$+2.525^{+0.033}_{-0.031}$ \\
   IH &$7.39^{+0.21}_{-0.20}$ &$-2.512^{+0.034}_{-0.031}$ \\
   \hline \hline
  \end{tabular}
  \caption{\small Neutrino mass data taken from
    Ref.\,\cite{Esteban:2018azc}, adopting data with the atmospheric
    neutrino by Super-Kamiokande. $\Delta m_{ij}^2\equiv m_i^2-m_j^2$,
    and $\Delta m_{3l}^2=\Delta m_{31}^2>0$ for the normal hierarchy
    (NH) case and $\Delta m_{3l}^2=\Delta m_{32}^2<0$ for the inverted
    hierarchy (IH) case.}
  \label{table:m_nu}
 \end{center}
\end{table}

Now let us discuss $M_\nu$. It can be diagonalized by a unitary matrix
$U_\nu$ as
\begin{align}
  D_{M_\nu}
  ={\rm diag}(m_1,m_2,m_3)=U_\nu^T M_\nu U_\nu\,.
\end{align}
As noted above, one of the three light neutrino masses is zero. We
follow the standard convention that $m_3>m_2>m_1(=0)$ for the normal
hierarchy (NH) case and $m_2>m_1>m_3(=0)$ for the inverted hierarchy
(IH) case and use the values given in Ref.\,\cite{Esteban:2018azc},
which are listed in Table~\ref{table:m_nu}.

Before discussing the parametrization of the neutrino Yukawa
couplings, it is instructive to count the number of parameters.  The
situation is the same as one discussed
Refs.\,\cite{Nakayama:2016gvg,Bjorkeroth:2016qsk} since the mass
matrix of the light neutrinos \eqref{eq:Mnu_lim} is similar.  Since
one neutrino is massless, there are 7 parameters in low energy, {\it
  i.e.}, 2 neutrino masses $+$ 3 real mixing angles $+$ 2 phases. On
the other hand, $M_\nu$ includes $y_{\nu ki}$ and $M_k$ where $k=1,2$
and $i=1,2,3$, which means 12 real parameters (neutrino Yukawa
couplings) $+$ 2 real parameters (right-handed neutrino
masses). However, 3 phases can be absorbed by lepton doublets and 2
real parameters are unphysical since $M_\nu$ is unchanged by the
rescalings $y_{\nu ki}\to \gamma_k y_{\nu ki}$ and $M_k \to \gamma^2_k
M_k$ with $\gamma_k$ being real constants. Therefore, we have 9
independent parameters in $M_\nu$ to determine 7 parameters in the
light neutrino sector. As it will be seen below, however, the
parametrization of the Yukawa couplings is different, especially for
$y_{\nu 3i}$ that are important parameters for the estimation of the
reheating temperature.

Let us discuss the NH case first.  We define $4\times 3$ matrix $R$ in
the similar manner in Refs.\,\cite{Casas:2001sr,Davidson:2002qv},
\begin{align}
  iR=D_{\tilde{M}}^{-1/2}U_{\tilde{M}}^T\tilde{m}_\nu
  U_\nu D_{M_\nu}^{-1/2}\,,
  \label{eq:R_NH}
\end{align}
which satisfies
\begin{align}
  R^TR={\rm diag}(0,1,1)\,.
  \label{eq:RTR}
\end{align}
Here $D_{\tilde{M}}^{-1/2}$ and $D_{M_\nu}^{-1/2}$ are matrices that
satisfy $(D_{\tilde{M}}^{-1/2})^2=D_{\tilde{M}}^{-1}$ and
$(D_{M_\nu}^{-1/2})^2={\rm diag}(0,m_2^{-1},m_3^{-1})$,
respectively. It is found that $R$ is more restrictive than
Eq.\,\eqref{eq:RTR}. Namely,
\begin{align}
  r^Tr=rr^T={\bf 1}\,,
  \label{eq:cond_for_R_NH}\\
  R_{42}/R_{32}=R_{43}/R_{33}=i\,, \\
  R_{l1}=0~~~~(l=1\mathchar`-\mathchar`-4)\,,
\end{align}
where
\begin{align}
  r\equiv \left(
\begin{array}{cc}
  R_{12} & R_{13} \\
  R_{22} & R_{23}  
\end{array} 
\right)\,.
\end{align}
Using the relations, the neutrino Yukawa couplings can be expressed in
terms of $R$. For later discussion, it is useful to give following
quantities:
\begin{align}
  (y_\nu y_\nu^\dagger)_{ii}&=M_i\sum_{j=2}^3 |R_{ij}|^2 m_j
  /\langle H_u^0 \rangle^2
  ~~~~~~~~~(i=1,2)\,,
  \label{eq:yydg_ii_NH}\\
  (y_\nu y_\nu^\dagger)_{33}&=2m_\phi\sum_{j=2}^3|R_{3j}|^2 m_j
   /\langle H_u^0 \rangle^2\,,
   \label{eq:yydg_33_NH}\\
   {\rm Im}\left[(y_\nu y_\nu^\dagger)_{21}^2\right]\frac{M_1}{M_2}&=
   -\frac{M_1^2}{\langle H_u^0\rangle^4}
        {\rm Im}\left[\sum_{j=2}^3R_{1j}^2m_j^2\right]\,,
        \label{eq:yy_21_NH} \\
        \sum_{i=1}^2{\rm Im}
        \left[(y_\nu y_\nu^\dagger)_{i3}^2\right]\frac{M_3}{M_i}&=
  -\frac{2M_3 m_\phi}{\langle H_u^0\rangle^4}
  {\rm Im}\left[\sum_{j=2}^3R_{3j}^2m_j^2\right]\,.
  \label{eq:yy_23_NH}
\end{align}
It is seen that $(y_\nu y_\nu^\dagger)_{11}$ and $(y_\nu
y_\nu^\dagger)_{22}$ are constrained by the neutrino oscillation data,
meanwhile $(y_\nu y_\nu^\dagger)_{33}$ is basically a free parameter
since $R_{3j}$ is not constrained. This is consistent with the fact
that $M_\nu$ is independent of $y_{\nu 3i}$.

The discussion is quite similar in the IH case.  The definition of $R$ is
the same form as in Eq.\,\eqref{eq:R_NH}, but satisfies $R^TR={\rm
  diag}(1,1,0)$ with $(D_{M_\nu}^{-1/2})^2={\rm
  diag}(m_1^{-1},m_2^{-1},0)$. Then, defining $r$ as
\begin{align}
  r\equiv \left(
\begin{array}{cc}
  R_{11} & R_{12} \\
  R_{21} & R_{22}  
\end{array} 
\right)\,,
\end{align}
we get
\begin{align}
  r^Tr=rr^T={\bf 1}\,,
  \label{eq:cond_for_R_IH}\\
  R_{41}/R_{31}=R_{42}/R_{32}=i\,, \\
  R_{l3}=0~~~~(l=1\mathchar`-\mathchar`-4)\,,
\end{align}
and the neutrino Yukawa couplings are given by,
\begin{align}
  (y_\nu y_\nu^\dagger)_{ii}&=M_i\sum_{j=1}^2 |R_{ij}|^2 m_j
  /\langle H_u^0 \rangle^2
  ~~~~~~~~~(i=1,2)\,,
  \label{eq:yydg_ii_IH}\\
  (y_\nu y_\nu^\dagger)_{33}&=2m_\phi\sum_{j=1}^2|R_{3j}|^2 m_j
  /\langle H_u^0 \rangle^2\,,
   \label{eq:yydg_33_IH}\\
   {\rm Im}\left[(y_\nu y_\nu^\dagger)_{21}^2\right]\frac{M_1}{M_2}&=
   -\frac{M_1^2}{\langle H_u^0\rangle^4}
        {\rm Im}\left[\sum_{j=1}^2R_{1j}^2m_j^2\right]\,,
        \label{eq:yy_21_IH} \\
        \sum_{i=1}^{2}
            {\rm Im}
            \left[(y_\nu y_\nu^\dagger)_{i3}^2\right]\frac{M_3}{M_i}&=
  -\frac{2M_3 m_\phi}{\langle H_u^0\rangle^4}
  {\rm Im}\left[\sum_{j=1}^2R_{3j}^2m_j^2\right]\,.
  \label{eq:yy_23_IH}
\end{align}

\section{Post inflationary regime}
\label{sec:post_inf}
\setcounter{equation}{0}

After the end of inflation, the inflaton oscillates around the global
minimum and decays eventually. Due to the decay the universe is
reheated and thermal plasma is created. In this section, we estimate
the reheating temperature and discuss how the lepton number asymmetry
is generated. As in the previous section, we take $\Omega\simeq
1$. After the universe is reheated, gravitinos are produced in various
ways. We discuss the gravitino problem at the end of this section.

\subsection{Reheating}
\label{sec:reheating}

The reheating temperature $T_R$ due to the inflaton decay is estimated
by,
\begin{align}
  T_R\simeq (90/\pi^2g_*(T_R))^{1/4}\sqrt{\Gamma_\phi M_{pl}}\,,
\end{align}
where $g_*(T)$ is the effective degree of freedom of radiation fields
at temperature $T$ and $\Gamma_\phi$ is the decay rate of the
inflaton. This expression is valid when the neutrino Yukawa couplings
that are responsible for the decay is sufficiently small to satisfy
$T_R\lesssim m_\phi$~\cite{Mukaida:2012qn,Mukaida:2012bz}, which is
the situation we focus on.\footnote{Of course, it is possible to
  consider a higher reheating temperature than the inflaton mass. Such
  a case is discussed in Ref.~\cite{Nakayama:2016gvg}. We will comment
  on the impact of such high reheating temperature on leptogenesis in
  the next subsection. } The inflaton decays as $\phi \to
L\tilde{H}_u$, $\bar{L}\bar{\tilde{H}}_u$, $\tilde{L}H_u$,
$\tilde{L}^*H_u^*$.\footnote{In general, the inflaton decays to
  gravitino pair or gravitino and right-handed neutrino. We will
  discuss those processes in Sec.\,\ref{sec:gravitino_problem}.}  Here
flavor indices and SU(2) doublet components are summed implicitly.
Then the decay rates for the modes are given by
\begin{align}
  &\Gamma_{\phi \to L\tilde{H}_u}=
  \Gamma_{\phi \to \bar{L}\bar{\tilde{H}}_u}=
  \frac{(y_{\nu}y_\nu^\dagger)_{33}}{16\pi}m_\phi\,, \\
 & \Gamma_{\phi \to \tilde{L}H_u}=
  \Gamma_{\phi \to \tilde{L}^*H_u^*}=
  \frac{(y_{\nu}y_\nu^\dagger)_{33}}{16\pi}\frac{M_3^2}{m_\phi}\,.
\end{align}
Since $\Gamma_{\phi \to \tilde{L}H_u}$ (=$\Gamma_{\phi \to
  \tilde{L}^*H_u^*}$) is suppressed by $(M_3/m_\phi)^2$, the total decay
rate is given by
\begin{align}
  \Gamma_\phi\simeq
  \Gamma_{\phi \to L\tilde{H}_u}+
  \Gamma_{\phi \to \bar{L}\bar{\tilde{H}}_u}=
  \frac{(y_{\nu}y_\nu^\dagger)_{33}}{8\pi}m_\phi\,.
  \label{eq:Gamma_phi}
\end{align}
Then the reheating temperature is estimated as
\begin{align}
  T_R\simeq 1.4\times 10^{10}\,{\rm GeV}
  \left(\frac{m_\phi}{10^{13}\,{\rm GeV}}\right)^{1/2}
  \left(\frac{(y_{\nu}y_\nu^\dagger)_{33}}{10^{-9}}\right)^{1/2}
  \left(\frac{g_{*}(T_R)}{228.75}\right)^{-1/4}\,.
  \label{eq:TR}
\end{align}
Recall that $(y_\nu y_\nu^\dagger)_{33}$ is not constrained by the
neutrino observations.  As a consequence, it is possible to
consider a wide range of values for the reheating temperature, which
is suitable for leptogenesis.

To end this subsection, we derive the number of $e$-folds before the
end of inflation. In this model, the inflaton oscillates after the end
of inflation and eventually decays to reheat the universe. Therefore,
it is given by
\begin{align}
  N_e\simeq&\, 55
  +\log\left(\frac{L}{{\rm Gpc}}\right)
  +\frac{1}{3}\log\left(\frac{T_R}{10^{10}\,{\rm GeV}}\right)
  \nonumber \\
  &+\log\left(\frac{H_e}{10^{13}\,{\rm GeV}}\right)
  -\frac{2}{3}\log\left(\frac{H_{\rm end}}{10^{13}\,{\rm GeV}}\right)\,,
\end{align}
where $L$ is the present cosmological scale, and $H_e$ and $H_{\rm
  end}$ are the Hubble scale corresponding to $N_e$ and at the end of
inflation, respectively. It is now clear that the parameters given in
Eq.\,\eqref{eq:parameter_space} is consistent with $N_e=55$\,--\,60.

\subsection{Leptogenesis}
\label{sec:leptogenesis}

Now we discuss the lepton number asymmetry. The lepton number is
generated via leptogenesis~\cite{Fukugita:1986hr} (see, for example,
Refs.\,\cite{Buchmuller:2005eh,Davidson:2008bu} for review). In the
following numerical study, we discuss following representative cases:
\begin{align}
  ({\rm I}).~&  M_1,M_2 < m_\phi  \,,
  \label{eq:caseI'}\\
  ({\rm II}).~& M_1,M_2 > m_\phi   \,.
  \label{eq:caseII'}
\end{align}
In both cases, $M_3$ should satisfy
Eq.\,\eqref{eq:constraint_on_Mi3}.

Let us consider case (I) first.  For simplicity, we consider the
Majorana masses are further hierarchical, {\it i.e.}, $M_1\ll
M_2$.\footnote{$M_3$ is irrelevant for leptogenesis if
  Eq.\eqref{eq:constraint_on_Mi3} is satisfied.  } Similar situation
has been studied intensively in the
literature~\cite{Giudice:2003jh,Buchmuller:2004nz}. Even though the
lepton number is generated by the inflaton decay, it is possibly
washed out when the reheating temperature is comparable or higher than
$M_{1}$.\footnote{It would be possible that $\tilde{N}_i$ ($i=1,2$)
  have initial amplitude of the Hubble parameter during inflation
  $H_{\rm inf}\sim g\xi/\sqrt{6}M_{pl}$. Then $\tilde{N}_i$ start to
  oscillate when $H\sim M_i$, and eventually decays. However, the
  effect of coherent oscillation of $\tilde{N}_i$ is negligible since
  the energy density ratio of $\tilde{N}_i$ to radiation at the decay
  is estimated as less than $\xi^2/18M_{pl}^4\sim 10^{-9}$.} To see
this more explicitly, it is convenient to introduce the effective
neutrino mass~\cite{Plumacher:1996kc} and equilibrium neutrino
mass~\cite{KolbTurner},
\begin{align}
  &\tilde{m}_1=\frac{(m_\nu m_\nu^\dagger)_{11}}{M_1}\,,
  \label{eq:mt1}\\
  &m_*=\frac{4\pi^2\sqrt{g_*(M_1)} \langle H_u^0\rangle^2 }{3\sqrt{10} M_{pl}}
  \simeq 3.9 \times 10^{-4}\,{\rm eV}
  \left(\frac{\langle H_u^0\rangle}{v/2}\right)^2\,,
  \label{eq:m_*}
\end{align}
where $v\simeq 246.7~{\rm GeV}$. If $\tilde{m}_1/m_*$ is larger than
unity, then it is the strong washout regime and the lepton number
generated at the reheating is washed out. $\tilde{m}_1$ is estimated
by using Eqs.\,\eqref{eq:yydg_ii_NH} and \eqref{eq:yydg_ii_IH},
\begin{align}
  \tilde{m}_1=
\left\{
  \begin{array}{ll}
   \sum_{j=2}^3 |R_{1j}|^2 m_j &({\rm NH}) \\
   \sum_{j=1}^2 |R_{1j}|^2 m_j  &({\rm IH})
   \end{array}\right.\,.
\end{align}
Using the neutrino mass data and Eqs.\,\eqref{eq:cond_for_R_NH} and
\eqref{eq:cond_for_R_IH}, it is straightforward to find that
$\tilde{m}_1$ has a lower bound,
\begin{align}
  \tilde{m}_1 \ge
  \left\{
  \begin{array}{ll}
   m_2 \simeq 8.6 \times 10^{-3}\,{\rm eV} &({\rm NH}) \\
   m_1 \simeq 4.9\times 10^{-2}\,{\rm eV}  &({\rm IH})
  \end{array}\right.\,.
  \label{eq:m1t_low}
\end{align}
Therefore, it is the strong washout regime in either case.

\begin{figure}[t]
  \begin{center}
    \includegraphics[scale=0.6]{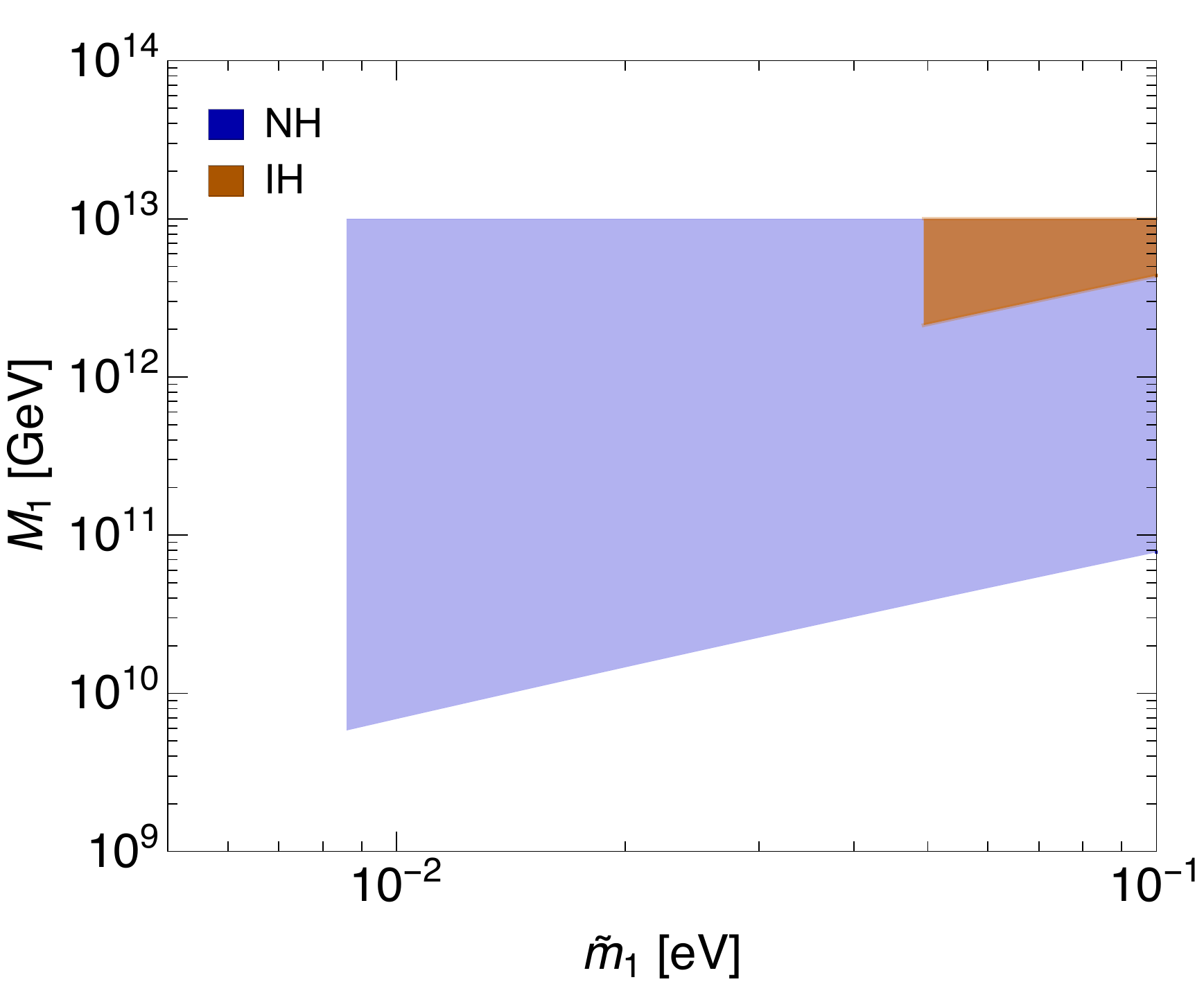}
  \end{center}
  \caption{Allowed region for $M_1$ as function of effective neutrino
    mass $\tilde{m}_1$ defined in Eq.\,\eqref{eq:mt1} for the normal
    hierarchy (NH) and inverted hierarchy (IH) cases. Lower bound on
    $M_1$ is obtained from $\eta_B^{\rm max}\ge \eta_B^{\rm obs}$
    while upper bound is \eqref{eq:caseI'}. Lower bound on
    $\tilde{m}_1$ is given by Eq.\,\eqref{eq:m1t_low}.}
  \label{fig:bound_for_M1}
\end{figure}

Although the primordial lepton number is washed out, the lepton number
is regenerated by the decay of the lightest right-handed (s)neutrino,
{\it i.e.}, $N_1$ and $\tilde{N}_1$ in the present case.  Then the
lepton number, strictly speaking lepton number minus baryon number, is
converted to baryon number via the sphaleron effect. This scenario
works if $T_R\gtrsim M_1$~\cite{Giudice:2003jh,Buchmuller:2004nz},
which is always possible as confirmed in the previous subsection. Then the
resultant baryon number becomes independent of $T_R$. In our study, we
adopt the analytic expressions in Ref.\,\cite{Buchmuller:2004nz} for
the calculation of the baryon number. Note that although the results
there are given in non-supersymmetric model, the results in
supersymmetric model do not change much both quantitatively and
qualitatively~\cite{Plumacher:1997ru,Campbell:1992hd,Davidson:2008bu}. In
our study we adopt the discussion given in
Ref.\,\cite{Davidson:2008bu}.  Then the baryon number is determined by
\begin{align}
  \eta_B\equiv \frac{n_B}{n_\gamma}=
  \frac{3\sqrt{2}}{4}\frac{a_{\rm sph}}{f}\epsilon_1 \kappa_f
  \simeq
  2.7\times 10^{-10} \left(\frac{\epsilon_1}{10^{-6}}\right)
  \left(\frac{\kappa_f}{2\times 10^{-2}}\right)\,,
  \label{eq:etaB}
\end{align}
where $n_B$ and $n_\gamma$ are number densities of baryon and photon at
present, respectively, $a_{\rm sph}=28/79$, $f=2387/86$, and a factor of
$\sqrt{2}$ counts the supersymmetric effect. The efficiency factor
$\kappa_f$ is given by~\cite{Buchmuller:2004nz} 
\begin{align}
  \kappa_f=(2\pm 1)\times 10^{-2}
  \left(\frac{0.01~{\rm eV}}{\tilde{m}_1}\right)^{1.1\pm 0.1}\,.
  \label{eq:kappa}
\end{align}
Finally, referring Ref.\,\cite{Covi:1996wh}, the asymmetric parameter
$\epsilon_1$ in our model is given by
\begin{align}
  \epsilon_1=
   -\frac{3}{16\pi}\frac{{\rm Im}\,\left[(y_\nu y_\nu^\dagger)_{21}^2\right] }
        {(y_\nu y_\nu^\dagger)_{11}}\frac{M_1}{M_2}
        =
  \frac{3}{16\pi}
  \frac{M_1}{\langle H^0_u \rangle^2}m_{\rm eff}\,,
\end{align}
where we have used Eqs.\eqref{eq:yydg_ii_NH}, \eqref{eq:yy_21_NH},
\eqref{eq:yydg_ii_IH}, and \eqref{eq:yy_21_IH} to obtain
\begin{align}
m_{\rm eff} =
\left\{
  \begin{array}{ll}
    \frac{{\rm Im}\sum_{j=2}^3R_{1j}^2m_j^2}{\sum_{j=2}^3|R_{1j}|^2m_j}
    &({\rm NH}) \\[4mm]
    \frac{{\rm Im}\sum_{j=1}^2R_{1j}^2m_j^2}{\sum_{j=1}^2|R_{1j}|^2m_j}
    &({\rm IH})
   \end{array}\right.\,.
\end{align}
It turns out that the maximum value of $m_{\rm eff}$, denoted as
$m_{\rm eff}^{\rm max}$, is
\begin{align}
  m_{\rm eff}^{\rm max} =
  \left\{
  \begin{array}{ll}
   m_3 - m_2 \simeq 4.2\times 10^{-2}\,{\rm eV}
    &({\rm NH}) \\
    m_2 - m_1\simeq 7.4 \times 10^{-4}\,{\rm eV}
    &({\rm IH})
   \end{array}\right.\,.
\end{align}
Therefore, parametrizing $m_{\rm eff}$
as $m_{\rm eff}=m_{\rm eff}^{\rm max} \sin \delta$, the asymmetric
parameter is given by
\begin{align}
  \epsilon_1\simeq
  \left\{
  \begin{array}{ll}
    8.2 \times 10^{-7}
    &({\rm NH}) \\
    1.5\times 10^{-8}
    &({\rm IH})
  \end{array}\right\}
  \times
  \left(\frac{M_1}{10^{10}\,{\rm GeV}}\right)
  \left(\frac{\langle H^0_u \rangle}{v/2}\right)^{-2}
  \left(\frac{\sin \delta}{0.5}\right)\,.
  \label{eq:epsilon_1}
\end{align}

Using Eqs.\,\eqref{eq:etaB}, \eqref{eq:kappa} and
\eqref{eq:epsilon_1}, the lower limit on $M_1$ to explain the present
baryon number is determined from $\eta_B^{\rm max} \ge \eta_B^{\rm
  obs}$ where $\sin \delta=1$ and $\eta_B^{\rm obs}$ is given
by~\cite{Aghanim:2018eyx}
\begin{align}
  \eta_B^{\rm obs}=
  (6.12\pm0.03)\times 10^{-10}\,.
\end{align}
In Fig.\,\ref{fig:bound_for_M1}, allowed regions are depicted for the
NH and IH cases. Here we consider so-called high-scale SUSY and take
$\langle H^0_u \rangle=v/2$ to get 125~GeV Higgs
mass~\cite{Giudice:2004tc,Giudice:2011cg}.  In the plot upper bound on
$M_1$ is given by \eqref{eq:caseI'}, {\it i.e.},
$M_1<m_\phi=10^{13}\,$GeV, and the lower bound on $\tilde{m}_1$ is
from Eq.\,\eqref{eq:m1t_low}.\footnote{Strictly speaking, the equality
  should be excluded since baryon number is zero.} The theoretical
uncertainties in Eq.\,\eqref{eq:kappa} are taken into account. It is
found that the present baryon number can be explained in a wide range
of parameter space for the NH case. For the IH case, on the other
hand, it seems that the parameter space for a successful leptogenesis
is relatively limited.  The lowest value required for $M_1$ turns out
to be
\begin{align}
  M_1 \gtrsim
   \left\{
  \begin{array}{ll}
    5.8 \times 10^{9}\,{\rm GeV}
    &({\rm NH}) \\
    2.1\times 10^{12}\,{\rm GeV}
    &({\rm IH})
  \end{array}\right. \,.
\end{align}
The lower limit is near the upper bound in the IH case. Here recall
that the upper bound on $M_1$ is just a theoretical one.  When $M_1\sim
m_\phi$, $T_R$ should be comparable to $m_\phi$, which is possible as
discussed in
Refs.\,\cite{Nakayama:2016gvg,Mukaida:2012qn,Mukaida:2012bz}.  In such
a case, sneutrino inflation and leptogenesis can be another source for
lepton asymmetry, which will be discussed below in detail. Therefore,
the upper bound merely indicates the parameter space for simple
thermal leptogenesis to work.

Let us move on to case (II). Since they are much heavier than the
inflaton, $N_{1,2}$ and $\tilde{N}_{1,2}$ are never thermalized after
the reheating.  For $N_3$ and $\tilde{N_3}$, on the other hand, it depends
on the effective neutrino mass that is defined by
\begin{align}
  \tilde{m}_3=\frac{(m_\nu m_\nu^\dagger)_{33}}{m_\phi}
  \simeq 1.5\times 10^{-9}\,{\rm eV}
  \left(\frac{(y_{\nu}y_{\nu}^\dagger)_{33}}{10^{-9}}\right)
  \left(\frac{10^{13}\,{\rm GeV}}{m_\phi}\right)
  \left(\frac{\langle H_u^0\rangle}{v/2}\right)^2
  \,.
  \label{eq:m3}
\end{align}
Then, from Eq.\,\eqref{eq:TR}, one obtains
\begin{align}
  \frac{T_R}{m_\phi}\simeq 0.71 \times
  \left(\frac{\tilde{m}_3}{m_*}\right)^{1/2}\,.
\end{align}
Here $m_*$ is defined similarly to Eq.\,\eqref{eq:m_*} but replacing
$g_*(M_1)$ by $g_*(m_\phi)$, and we have taken $g_*(T_R)\simeq
g_*(m_\phi)$.  As explained in Sec.\,\ref{sec:reheating}, the
expression Eq.\,\eqref{eq:TR} is valid for $T_R/m_\phi \lesssim 1$
that is satisfied for $\tilde{m}_3<m_*$. Such a case corresponds to
the weak washout regime. In that regime, the $N_3$ and $\tilde{N_3}$
are not thermalized, and the lepton number produced by the inflaton
decay can be the source of the present baryon number.  This situation
is similar to sneutrino inflation and
leptogenesis~\cite{Murayama:1992ua,Murayama:1993xu,Murayama:1993em,Hamaguchi:2001gw,Ellis:2003sq,Antusch:2004hd,Antusch:2009ty,Kadota:2005mt,Nakayama:2013nya,Nakayama:2016gvg,Bjorkeroth:2016qsk}. (See
also Refs.~\cite{Binetruy:1987xj,Gaillard:1995az,Evans:2015mta} for
leptogenesis via Afflec-Dine mechanism~\cite{Affleck:1984fy}.)

\begin{figure}[t]
  \begin{center}
    \includegraphics[scale=0.6]{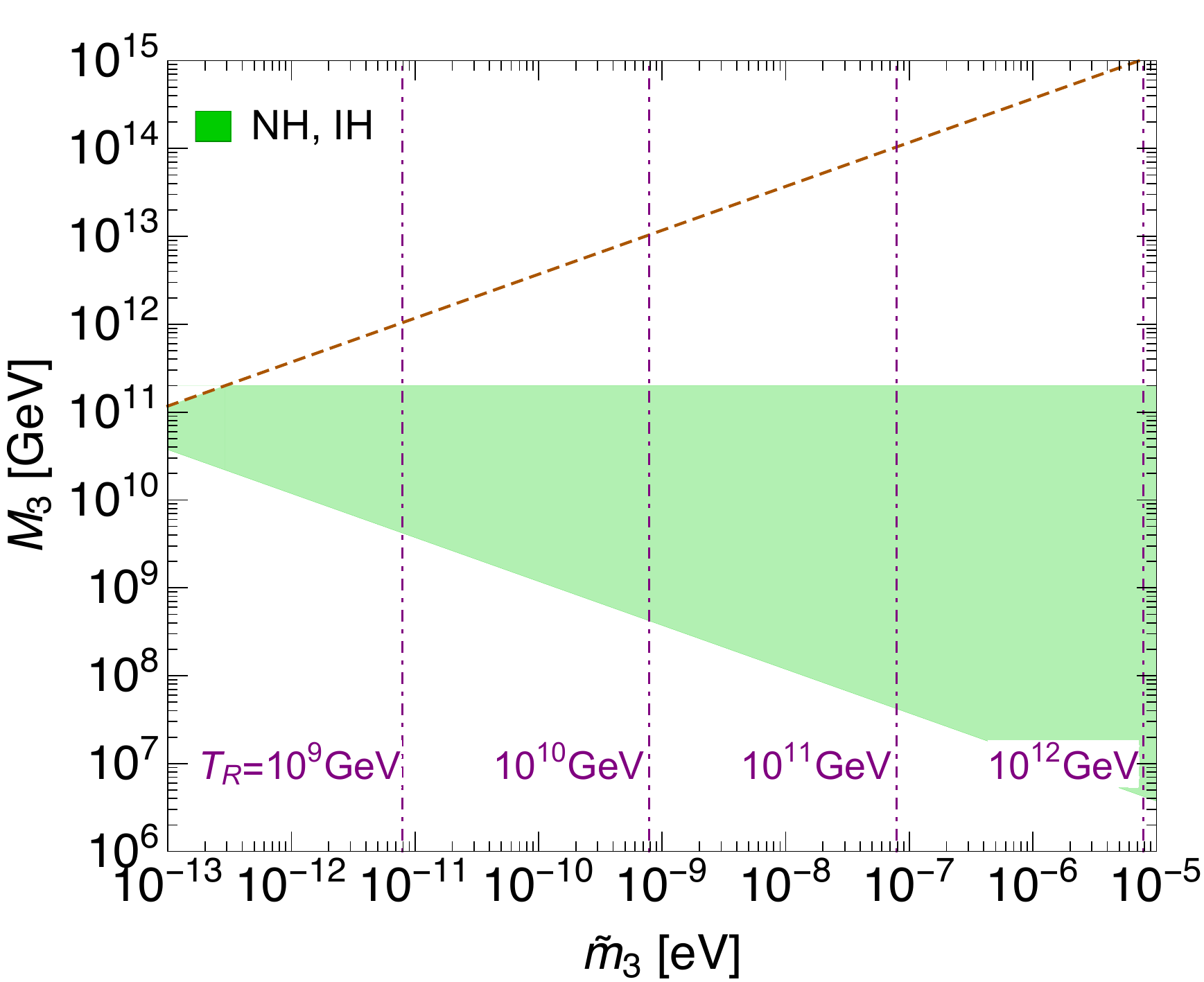}
  \end{center}
  \caption{Allowed region for $M_3$ as function of $\tilde{m}_3$
    defined in Eq.\,\eqref{eq:m3}. Results are the same for the NH and
    IH cases. Lower bound on $M_3$ comes from $\eta_B^{\rm max}\ge
    \eta_B^{\rm obs}$ meanwhile upper one is given by
    Eqs.\,\eqref{eq:constraint_on_Mi3} and \eqref{eq:stability}.  The
    former is not to affect the inflationary trajectory and the latter
    (dashed orange) is from the stability of the scalar
    potential. Vertical lines (dot-dashed purple) are contours of
    $T_R$.}
  \label{fig:bound_for_M3}
\end{figure}

Let us suppose $\tilde{m}_3\ll m_*$, {\it i.e.}, $T_R/m_\phi \ll 1$.
Then baryon number is given by
\begin{align}
  \eta_B=\frac{3}{4}\frac{T_R}{m_\phi}a_{\rm sph}^{\rm MSSM}d\epsilon_\phi\,,
\end{align}
where $a_{\rm sph}^{\rm MSSM}=8/23$ and
$d=(s/n_\gamma)_0=43\pi^4/495\zeta(3)$ is the present value of entropy
density to and photon density ratio. $\epsilon_\phi$ is obtained by
an explicit calculation as
\begin{align}
  \epsilon_\phi =
  -\frac{3}{4\pi}\sum_{i=1}^2
  \frac{{\rm Im}\,\left[(y_\nu y_\nu^\dagger)_{i3}^2\right] }
       {(y_\nu y_\nu^\dagger)_{33}}\frac{M_3}{M_i}
       =
\frac{3}{4\pi}\frac{M_3}{\langle H_u^0 \rangle^2}m'_{\rm eff}
\,,
\label{eq:epsilon_phi}
\end{align}
where
\begin{align}
m'_{\rm eff} =
\left\{
  \begin{array}{ll}
    \frac{{\rm Im}\sum_{j=2}^3R_{3j}^2m_j^2}
         {\sum_{j=2}^3|R_{3j}|^2m_j}
    &({\rm NH}) \\[4mm]
         \frac{{\rm Im}\sum_{j=1}^2R_{3j}^2m_j^2}
              {\sum_{j=1}^2|R_{3j}|^2m_j}
    &({\rm IH})
   \end{array}\right.\,.
\end{align}
In the second step, we have used Eqs.\,\eqref{eq:yydg_33_NH},
\eqref{eq:yy_23_NH}, \eqref{eq:yydg_33_IH}, and \eqref{eq:yy_23_IH}.
It should be noted that $\epsilon_\phi$ is independent of the inflaton
mass, but it depends on $M_3$.  Even though $R_{3j}$ are not
constrained, it has been found that $m'_{\rm eff}$ is bounded from
above. The maximum value turns out to be
\begin{align}
  m_{\rm eff}^{\prime\,\rm max} =
  \left\{
  \begin{array}{ll}
   m_3 \simeq 5.0\times 10^{-2}\,{\rm eV}
    &({\rm NH}) \\
    m_2\simeq 5.0 \times 10^{-2}\,{\rm eV}
    &({\rm IH})
   \end{array}\right.\,.
\end{align}
Therefore, parametrizing the effective neutrino mass as $m'_{\rm
  eff}=m_{\rm eff}^{\prime\,{\rm max}}\sin\delta'$ the asymmetric
parameter is given by
\begin{align}
  \epsilon_\phi\simeq
      3.9\times 10^{-9}
    \times
  \left(\frac{M_3}{10^{7}\,{\rm GeV}}\right)
  \left(\frac{\langle H^0_u \rangle}{v/2}\right)^{-2}
  \left(\frac{\sin \delta'}{0.5}\right)\,.
\end{align}
for both the NH and IH cases. Using the equations, we get
\begin{align}
  \eta_B^{\rm max}\simeq
  1.6\times 10^{-10}
  \left(\frac{M_3}{10^7\,{\rm GeV}}\right)
  \left(\frac{\tilde{m}_3}{10^{-7}\,{\rm eV}}\right)^{1/2}
  \left(\frac{\langle H_u^0 \rangle}{v/2}\right)^{-3}\,.
\end{align}
Since $\eta_B$ is independent of $m_\phi$, the requirement $\eta_B^{\rm
  max}\ge \eta_B^{\rm obs}$ gives a lower bound on $M_3$, which is
plotted in Fig.\,\ref{fig:bound_for_M3}. Here $\langle H^0_u
\rangle=v/2$ is taken as in Fig.\,\ref{fig:bound_for_M1}. Upper bound
$M_3<2\times 10^{11}\,{\rm GeV}$ is from
Eq.\eqref{eq:constraint_on_Mi3}.  Another upper bound on $M_3$ from
the stability of the inflationary trajectory, Eq.\,\eqref{eq:stability},
is also shown.  We quit plotting for region $\tilde{m}_3>10^{-5}\,{\rm
  eV}$ because $\tilde{m}_3\ll m_*$ is no longer valid. In the plot,
contours of $T_R$ are depicted. It is found that the leptogenesis is
successful in a wide parameter space for both the NH and IH cases.
Lower bound on $M_3$ behaves similarly to region C in Fig.\,1 of
Ref.~\cite{Ellis:2003sq} by reading $M_1$ and $\tilde{m}_1$ as $M_3$
and $\tilde{m}_3$, respectively, {\it i.e.}, the lower bound is
proportional to $1/\sqrt{\tilde{m}_3}$. Quantitatively, the lower
bound in our model is relaxed by roughly a factor of $4$ compared to
the result in the reference. This can be understood as follows; first,
the decay rate of the inflaton in our model is different from one in the
literature (see Eq.\,\eqref{eq:Gamma_phi}), leading to a $1/\sqrt{2}$
suppression in $T_R/m_\phi$; as another consequence, the expression
\eqref{eq:epsilon_phi} (with $m_{\rm eff}^{\prime}=m_{\rm eff}^{\prime
  {\rm max}}$) is enhanced by a factor of two compared to
$|\epsilon_1^{\rm max}|$ in the reference; finally $\tan \beta=\infty$
is taken in the work meanwhile $\tan \beta =1$ in our model. Thus, in
total, a factor of $4$ enhancement is obtained in $\eta_B$.

In the case where $\tilde{m}_3$ gets much larger than $m_*$, the
situation reduces to case (I). Namely, the reheating temperature is so
high that both $\tilde{N}_3$ and $N_3$ are thermalized and thermal
leptogenesis takes place. Resultant allowed region is the same as the
NH case of Fig.\,\ref{fig:bound_for_M1}, by replacing $M_1$ and
$\tilde{m}_1$ by $M_3$ and $\tilde{m}_3$, respectively, but there is
no lower bound on $\tilde{m}_3$ meanwhile there is the upper bound on
$M_3$.  In the intermediate case, $\tilde{m}_3\sim m_*$, on the other
hand, the Boltzmann equations should be solved numerically to get the
lepton number, which is already done in Ref.\,\cite{Ellis:2003sq}. The
result corresponds to region B in Fig.\,1 of the reference. Strictly
speaking, the effective dissipation rate should be used instead of the
decay rate of the inflaton~\cite{Mukaida:2012bz} in the Boltzmann
equations. As shown in the reference, the reheating process is so
efficient when the dissipation rate is taken into account that the
reheating temperature can exceed the mass of the inflaton mass and
consequently $N_3$ and $\tilde{N}_3$ are easily thermalized. Once they
are thermalized, the thermal leptogenesis takes place, where the
resultant baryon number becomes independent of the reheating
temperature. Eventually the situation reduces to the case (I). Such
qualitative behavior can be confirmed by numerical study, which is
left for the future work.

Crucial difference from sneutrino
leptogenesis~\cite{Murayama:1992ua,Murayama:1993xu,Murayama:1993em,Hamaguchi:2001gw,Ellis:2003sq}
is that although $M_3$ and $\tilde{m}_3$, {\it i.e.}, $(y_\nu
y_\nu^\dagger)_{33}$, are important parameters to determine baryon
number, they are sequestered from other physical quantities, such as
the heavy right-handed (s)neutrino masses or the light neutrino mass
matrix. Therefore, there is no consequence in other low energy
experiments.  This is a feature of case (II).

\subsection{Gravitino problem}
\label{sec:gravitino_problem}

In the framework of supergravity, a fair amount of gravitino
$\psi_\mu$ can be produced in various ways in the thermal history of
the universe. Since the interactions of gravitino with the MSSM
particles are Planck-suppressed, gravitino is long-lived and its decay
can spoil the successful big-bang nucleosynthesis if it is
unstable. Although this problem can be avoided when gravitino is
enough heavy to have the lifetime much shorter than 1 sec, gravitino
decay produces the lightest superparticle (LSP). Then the LSP produced
by the decay may overclose the universe if the R-parity is conserved.

There are three types of production mechanism of gravitino in the
model we consider; (i) the inflaton decay; (ii) thermal scattering
from the thermal
bath~\cite{Bolz:2000fu,Pradler:2006qh,Pradler:2006hh,Rychkov:2007uq};
(iii) decay of superparticles in the thermal
bath~\cite{Asaka:2005cn,Cheung:2011nn}.

In general, process (i) includes gravitino pair production. The decay
width of the mode, however, depends on the inflaton
VEV~\cite{Kawasaki:2006gs,Kawasaki:2006hm,Asaka:2006bv,Endo:2006tf,Endo:2007ih,Endo:2007sz}. In
our case, therefore, this process can be ignored since the inflaton
does not have a VEV. On the other hand, the inflaton can decay to
gravitino and right-handed neutrino. The decay width is given by
\begin{align}
  \Gamma_{\phi\to \psi_{\mu}N_3}=
  \frac{\beta_f^{3}m_\phi^5}{48\pi M_{pl}^2m_{3/2}^2}
  \left[1-\frac{(m_f+m_{3/2})^2}{m_\phi^2}\right]\,,
\end{align}
where
$\beta_f^{2}=1-2(m_f^2+m_{3/2}^2)/m_\phi^2+(m_f^2-m_{3/2}^2)^2/m_\phi^4\,.$
Here $m_f$ and $m_{3/2}$ are masses of $N_3$ and $\psi_\mu$,
respectively. The mass difference between $\phi$ and $N_3$ is expected
to be given by the soft SUSY breaking mass scale for scalar
superpartners, $|m_\phi-m_f|\sim \tilde{m}$.  Let us assume this decay
happens by taking $m_\phi=m_f+\tilde{m}$ where $\tilde{m}=k m_{3/2}$
($k>1$). In the limit $km_{3/2}/m_\phi\ll 1$, we obtain
\begin{align}
  \Gamma_{\phi\to \psi_{\mu}N_3}\simeq
  \frac{Cm_\phi m_{3/2}^2}{3\pi M^2_{pl}}\,,
\end{align}
where $C=(k-1)(k^2-1)^{3/2}$. Then the branching fraction of this mode is
\begin{align}
  {\rm Br}_{\phi\to \psi_{\mu}N_3} =
  \frac{\Gamma_{\phi\to \psi_{\mu}N_3}}{\Gamma_{\phi}}\simeq
  4.5\times 10^{-16}C\left(\frac{m_{3/2}}{10^{5}\,{\rm GeV}}\right)^2
  \left(\frac{10^{-11}}{(y_\nu y_\nu^\dagger)_{33}}\right)\,.
\end{align}
Therefore, it is sure that the inflaton decay reheats the universe in
wide range of gravitino mass region. On the other hand, the resultant
gravitino abundance produced by the decay is estimated as
\begin{align}
  \Omega_{3/2}^{\rm inf}h^2\sim
  1.3\times 10^{-6} C\left(\frac{m_{3/2}}{10^{5}\,{\rm GeV}}\right)^3
  \left(\frac{10^{13}\,{\rm GeV}}{m_{\phi}}\right)^{1/2}
  \left(\frac{10^{-11}}{(y_\nu y_\nu^\dagger)_{33}}\right)^{1/2}\,,
\end{align}
where $h$ is the scale factor of Hubble expansion rate.

Gravitino abundance via process (ii) is most effective at high
temperature, thus it is proportional to $T_R$, meanwhile in process
(iii) gravitino is dominantly produced when the temperature is around
the mass of decaying particle.  Adopting the expression given in
Ref.\,\cite{Hall:2012zp}, the abundances via processes (ii) and (iii)
are given by
\begin{align}
  &\Omega_{3/2}^{\rm TH}h^2\sim
  4
  \left(\frac{T_R}{10^{9}\,{\rm GeV}}\right)
  \left(\frac{m_{3/2}}{10^{5}\,{\rm GeV}}\right)\,,
  \\
  &\Omega_{3/2}^{\rm FI}h^2\sim
  5\times 10^{-4}k^3
  \left(\frac{m_{3/2}}{10^5\,{\rm GeV}}\right)^2\,.
  \label{eq:Omega_{3/2}^FI}
\end{align}
Here the contribution of the longitudinal mode of gravitino is
suppressed in $\Omega_{3/2}^{\rm TH}$ by considering gluino is lighter
than gravitino.  In $\Omega_{3/2}^{\rm FI}$, we have assumed that all
scalar leptons and quarks in the MSSM (whose mass scale is
$\tilde{m}$) are heavier than gravitino and that they are
thermalized. And as in the discussion of the inflaton decay,
$\tilde{m}=km_{3/2}$ has been taken.\footnote{We have checked that the
  contribution from $\tilde{N}_1$ is negligible even if $\tilde{N}_1$
  is thermalized, {\it i.e.}, in case (I).}  Then the relic abundance
of the LSP due to gravitino decay is estimated as
\begin{align}
  \Omega_{\rm LSP}^{\rm non\mathchar`- th}h^2
  =\frac{m_{\rm LSP}}{m_{3/2}}\Omega_{3/2}h^2\,,
\end{align}
where $m_{\rm LSP}$ is the LSP mass and
$\Omega_{3/2}=\Omega_{3/2}^{\rm inf}+\Omega_{3/2}^{\rm
  TH}+\Omega_{3/2}^{\rm FI}$ is the total gravitino
abundance. $\Omega_{\rm LSP}^{\rm non\mathchar`- th}h^2$ should not
exceed the observed dark matter abundance $\Omega_{\rm DM}h^2\simeq
0.12$~\cite{Aghanim:2018eyx}, which gives a constraint on gravitino
mass.

\begin{figure}[t]
  \begin{center}
    \includegraphics[scale=0.42]{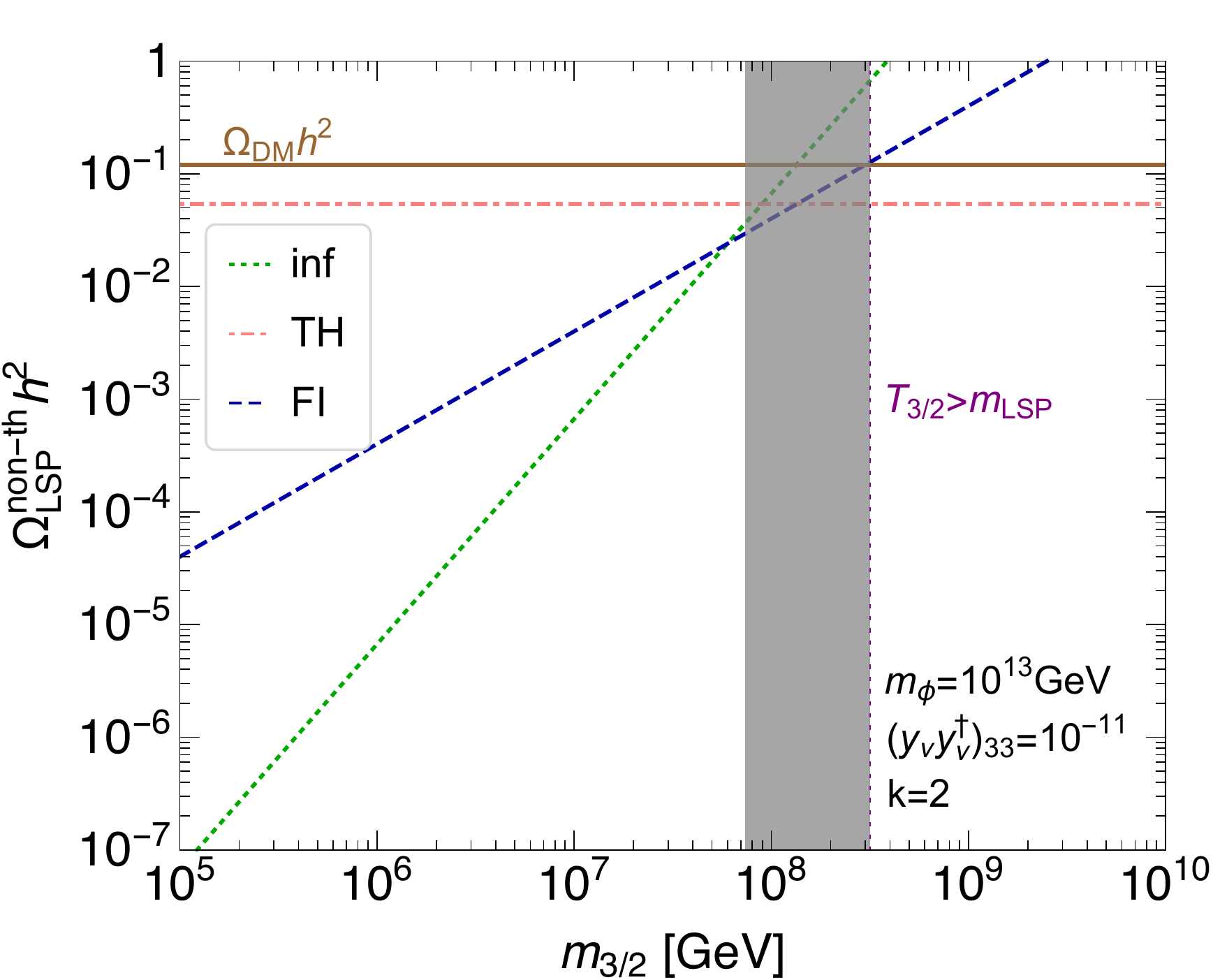}
    \includegraphics[scale=0.42]{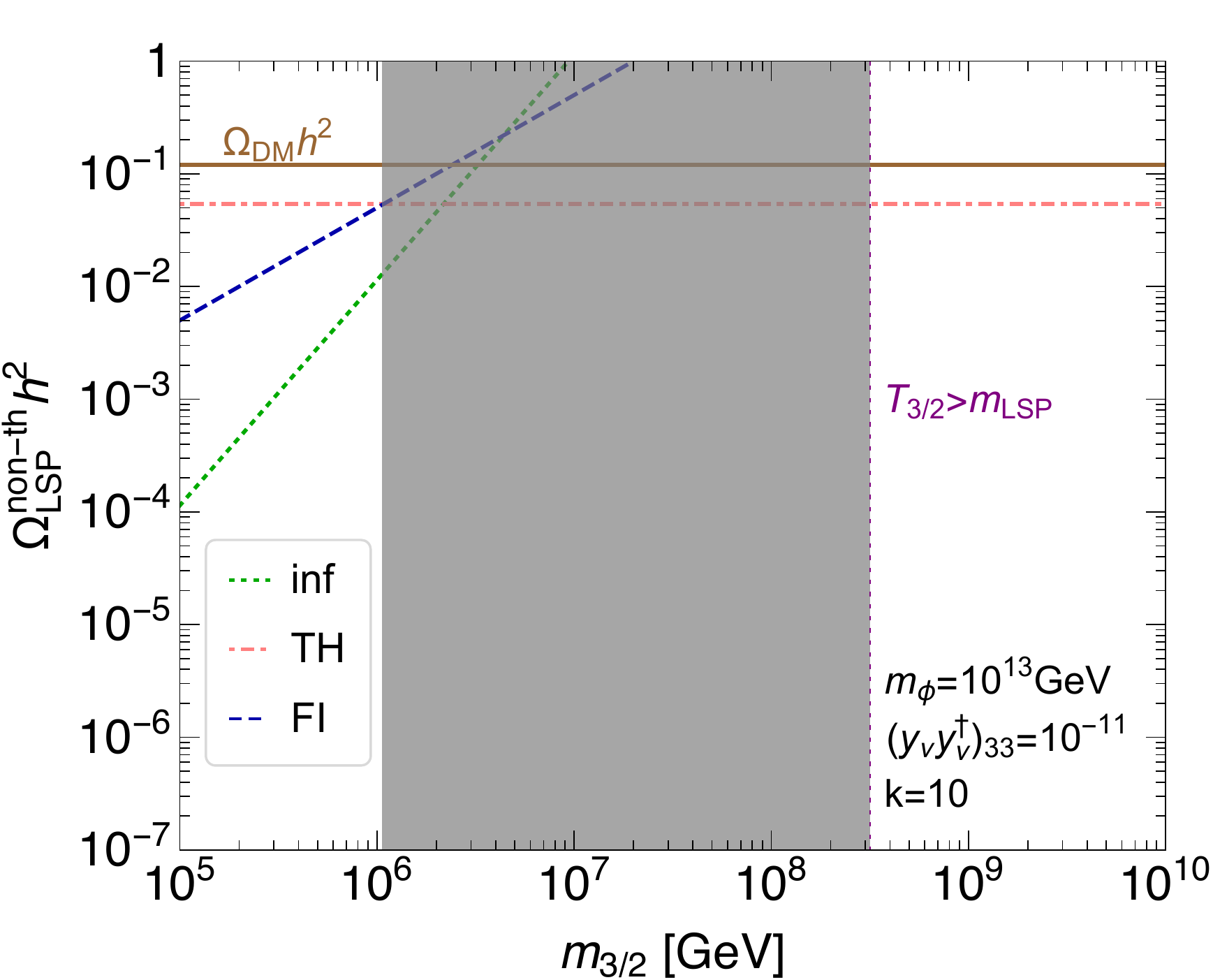}
  \end{center}
  \caption{Relic density of the LSP as function of gravitino mass.
    ``inf'' (dotted green), ``TH'' (dot-dashed red), and ``FI''
    (dashed blue) are the contributions from process (i), (ii), and
    (iii), respectively. $\Omega_{\rm DM}h^2\simeq 0.12$ (solid brown) is
    indicated as a reference. Right region from vertical line (dotted
    violet) indicates $T_{3/2}>m_{\rm LSP}$, and shaded region is
    excluded. $m_{\rm LSP}=1\,{\rm TeV}$, $m_\phi=10^{13}\,{\rm GeV}$,
    $(y_\nu y_\nu^\dagger)_{33}=10^{-11}$, and $k=\tilde{m}/m_{3/2}=2$
    (left), 10 (right) are taken.}
  \label{fig:omg}
\end{figure}

Fig.\,\ref{fig:omg} shows the resultant $\Omega_{\rm LSP}^{\rm
  non\mathchar`- th}$ from the each contribution.  Here $m_{\rm
  LSP}=1\,{\rm TeV}$ is taken by considering 1 TeV Higgsino or Wino
dark matter with a mass of 2.7\,--\,3 TeV~\cite{Hisano:2006nn}.
$m_\phi=10^{13}\,{\rm GeV}$, $(y_\nu y_\nu^\dagger)_{33}=10^{-11}$,
and $k=2$ (left), 10 (right) are taken to determine the contributions
from processes (i) and (iii). It is seen that in lower gravitino mass
region, the dominant contribution to $\Omega_{\rm LSP}^{\rm
  non\mathchar`- th}$ is from process (ii). In order for the
contribution not to exceed the dark matter abundance, $T_R\lesssim
10^9\,{\rm GeV}$ is required, which is well-known result. This gives a
stringent constraint on the parameter space for successful
leptogenesis shown in Figs.\,\ref{fig:bound_for_M1} and
\ref{fig:bound_for_M3} if gravitino is not heavy enough. Namely,
$\tilde{m}_1\sim 10^{-2}\,{\rm eV}$ is the allowed region in case (I)
meanwhile $10^{-13}\,{\rm eV}\lesssim \tilde{m}_3\lesssim
10^{-11}\,{\rm eV}$ is allowed in case (II).  On the other hand, the
contributions from processes (i) and (iii) depend on the parameters,
especially $k=\tilde{m}/m_{3/2}$ (and $m_{3/2}$). In order for
$\Omega_{\rm LSP}^{\rm non\mathchar`- th}$ not to exceed the dark
matter abundance, the upper bound on gravitino mass is obtained
depending on the mass spectrum of squarks and sleptons, {\it e.g.},
$m_{3/2}\lesssim 10^8$ ($10^{6}$)\,GeV for $\tilde{m}\sim m_{3/2}$
$(10m_{3/2})$. Such gravitino mass is preferred in minimal or mini
split supersymmetry~\cite{Arvanitaki:2012ps,ArkaniHamed:2012gw}, pure
gravity mediation~\cite{Ibe:2011aa,Ibe:2012hu}, and spread
supersymmetry~\cite{Hall:2012zp,Hall:2011jd}.

On the other hand, there is also an allowed region in higher gravitino
mass region. This is because gravitino decays before thermal
freeze-out of the LSP in that region. The allowed region can be
estimated by imposing gravitino decay temperature $T_{3/2}$ larger
than the LSP mass. The gravitino decay temperature is defined by
\begin{align}
  T_{3/2}\simeq (90/\pi^2g_*(T_{3/2}))^{1/4}\sqrt{\Gamma_{3/2} M_{pl}}\,,
\end{align}
where $\Gamma_{3/2}$ is the decay rate of gravitino. Then
$m_{3/2}\gtrsim 3\times 10^{8}\,{\rm GeV}$ is obtained from
$T_{3/2}>m_{\rm LSP}$ for $m_{\rm LSP}=1\,{\rm TeV}$ (see {\it e.g.},
Ref.\,\cite{Ishiwata:2013waa}). Such high gravitino mass can be
considered high-scale SUSY~\cite{Hall:2009nd}, intermediate scale
supersymmetry~\cite{Hall:2014vga}, and unified inflation
model~\cite{Domcke:2017rzu}. However, gravitino cannot be too heavy
because ${\rm Br}_{\phi\to \psi_{\mu}N_3}$ should be less than unity
for the reheating. Taking ${\rm Br}_{\phi\to \psi_{\mu}N_3} \lesssim
0.1$, for example, upper bound on gravitino mass is obtained as
$m_{3/2}\lesssim 5\times 10^{11}\,{\rm GeV}$ ($2\times 10^{10}\,{\rm
  GeV}$) for $k=2$ (10).

Another option is the R-parity violation. Under the R-parity
violation, the LSP decays to the standard-model particles. Then, the
LSP does not contribute to the matter abundance of the universe so
that there is no constraint on $m_{3/2}$.

\section{Conclusion}
\label{sec:conclusion}

Superconformal subcritical hybrid inflation is one of attractive
inflation models that are consistent with the observed cosmological
parameters by the Planck satellite. In this paper we have studied the
cosmology of an extended version of the model.  In the model three
right-handed neutrinos are introduced. The superpotential consists of
one in the supersymmetric seesaw model and the interaction terms of
the right-handed neutrinos with the additional matter fields, one of
which plays the role of the waterfall field.  In the K\"{a}hler
potential, on the other hand, it is possible for the sneutrinos to
have shift symmetry by introducing explicit superconformal breaking
terms of ${\cal O}(1)$. Due to the shift symmetry, one of the
sneutrinos becomes the inflaton field similarly in superconformal
subcritical hybrid inflation. Although the mass terms of the
sneutrinos can affect the trajectory of the inflaton, it has turned
out that the effect is restrictive and viable inflation is realized.
After inflation, the inflaton field decays to Higgses and sleptons or
Higgsinos and leptons to reheat the universe.

Light neutrino masses are given by the seesaw mechanism. However, the
mass matrix is different from the conventional one.  It turns out that
one of the neutrinos is massless.  Assuming that suppressed couplings
of the other right-handed neutrinos to the waterfall field, it has
been found that the neutrino Yukawa couplings that couples the
inflaton to the MSSM sector are not constrained by the neutrino
oscillation data. Consequently, the reheating temperature is a free
parameter, which is suitable for leptogenesis.

We have considered two representative cases; (I) the other
right-handed (s)neutrinos are lighter than the inflaton; (II) the other
right-handed (s)neutrinos are heavier than the inflaton.  In case (I),
thermal leptogenesis is possible if the reheating temperature is
larger than $\sim 10^{9}\,{\rm GeV}$.  It has been found leptogenesis
is successful in a wide range of parameter space in the normal
hierarchy case while the parameter space for leptogenesis is
relatively limited in the inverted hierarchy case. In case (II), on
the other hand, sneutrino leptogenesis takes place if the reheating
temperature is larger than $\sim 10^{8}\,{\rm GeV}$. It has turned out
that in both the normal and inverted hierarchy cases successful
leptogenesis is realized in wide range of parameter space.

\section*{Acknowledgments}
We are grateful to Wilfried~Buchm\"{u}ller, Pasquale~Di~Bari,
Oleg~Lebedev, and Fuminobu~Takahashi for valuable discussions.  This
work was supported by JSPS KAKENHI Grant Numbers JP17H05402,
JP17K14278, JP17H02875 and JP18H05542 (KI).

\appendix


\begin{thebibliography}{99}

\bibitem{Kawasaki:2000yn} 
  M.~Kawasaki, M.~Yamaguchi and T.~Yanagida,
  Phys.\ Rev.\ Lett.\  {\bf 85}, 3572 (2000)
  doi:10.1103/PhysRevLett.85.3572
  [hep-ph/0004243].

\bibitem{Buchmuller:2014rfa} 
  W.~Buchmuller, V.~Domcke and K.~Schmitz,
  JCAP {\bf 1411}, no. 11, 006 (2014)
  doi:10.1088/1475-7516/2014/11/006
  [arXiv:1406.6300 [hep-ph]].

\bibitem{Buchmuller:2014dda} 
  W.~Buchmuller and K.~Ishiwata,
  Phys.\ Rev.\ D {\bf 91}, no. 8, 081302 (2015)
  doi:10.1103/PhysRevD.91.081302
  [arXiv:1412.3764 [hep-ph]].
  
\bibitem{Buchmuller:2012ex} 
  W.~Buchmüller, V.~Domcke and K.~Schmitz,
  JCAP {\bf 1304}, 019 (2013)
  doi:10.1088/1475-7516/2013/04/019
  [arXiv:1210.4105 [hep-ph]].

\bibitem{Buchmuller:2013zfa} 
  W.~Buchmuller, V.~Domcke and K.~Kamada,
  Phys.\ Lett.\ B {\bf 726}, 467 (2013)
  doi:10.1016/j.physletb.2013.08.042
  [arXiv:1306.3471 [hep-th]].

\bibitem{Starobinsky:1980te} 
  A.~A.~Starobinsky,
  Phys.\ Lett.\ B {\bf 91}, 99 (1980)
  [Phys.\ Lett.\  {\bf 91B}, 99 (1980)]
  [Adv.\ Ser.\ Astrophys.\ Cosmol.\  {\bf 3}, 130 (1987)].
  doi:10.1016/0370-2693(80)90670-X
  
\bibitem{Einhorn:2009bh} 
  M.~B.~Einhorn and D.~R.~T.~Jones,
  JHEP {\bf 1003}, 026 (2010)
  doi:10.1007/JHEP03(2010)026
  [arXiv:0912.2718 [hep-ph]].

\bibitem{Kallosh:2010ug} 
  R.~Kallosh and A.~Linde,
  JCAP {\bf 1011}, 011 (2010)
  doi:10.1088/1475-7516/2010/11/011
  [arXiv:1008.3375 [hep-th]].

\bibitem{Ferrara:2010yw} 
  S.~Ferrara, R.~Kallosh, A.~Linde, A.~Marrani and A.~Van Proeyen,
  Phys.\ Rev.\ D {\bf 82}, 045003 (2010)
  doi:10.1103/PhysRevD.82.045003
  [arXiv:1004.0712 [hep-th]].

\bibitem{Ferrara:2010in} 
  S.~Ferrara, R.~Kallosh, A.~Linde, A.~Marrani and A.~Van Proeyen,
  Phys.\ Rev.\ D {\bf 83}, 025008 (2011)
  doi:10.1103/PhysRevD.83.025008
  [arXiv:1008.2942 [hep-th]].
  
\bibitem{Kallosh:2013yoa} 
  R.~Kallosh, A.~Linde and D.~Roest,
  JHEP {\bf 1311}, 198 (2013)
  doi:10.1007/JHEP11(2013)198
  [arXiv:1311.0472 [hep-th]].

\bibitem{Kallosh:2013xya} 
  R.~Kallosh and A.~Linde,
  JCAP {\bf 1306}, 028 (2013)
  doi:10.1088/1475-7516/2013/06/028
  [arXiv:1306.3214 [hep-th]].

\bibitem{Ishiwata:2018dxg} 
  K.~Ishiwata,
  Phys.\ Lett.\ B {\bf 782}, 367 (2018)
  doi:10.1016/j.physletb.2018.05.047
  [arXiv:1803.08274 [astro-ph.CO]].
  
\bibitem{Bryant:2016tzg} 
  B.~C.~Bryant and S.~Raby,
  Phys.\ Rev.\ D {\bf 93}, no. 9, 095003 (2016)
  doi:10.1103/PhysRevD.93.095003
  [arXiv:1601.03749 [hep-ph]].

\bibitem{Bryant:2016sjj} 
  B.~C.~Bryant, Z.~Poh and S.~Raby,
  arXiv:1612.04382 [hep-ph].

\bibitem{Domcke:2017rzu} 
  V.~Domcke and K.~Schmitz,
  Phys.\ Rev.\ D {\bf 97}, no. 11, 115025 (2018)
  doi:10.1103/PhysRevD.97.115025
  [arXiv:1712.08121 [hep-ph]].
  
\bibitem{seesaw}
  P.~Minkowski,
  Phys.\ Lett.\  {\bf 67B}, 421 (1977)
  doi:10.1016/0370-2693(77)90435-X;
  T.~Yanagida,
  in “Proceedings of the Workshop on Unified Theory and
  Baryon Number of the Universe,” Tsukuba, Japan, Feb.\ 13-14, 1979,
  p.\ 95, eds. O.~Sawada and A.~Sugamoto (KEK Report KEK-79-18, 1979,
  Tsukuba);
  Prog.\ Theor.\ Phys.\  {\bf 64}, 1103 (1980)
  doi:10.1143/PTP.64.1103;
  M.~Gell-Mann, P.~Ramond and R.~Slansky, in “Supergravity,”
  eds. P.~van~Niewwenhuizen and D.~Freedman (North Holland,Amsterdam
  1980); P.~Ramond, in Talk given at the Sanibel Symposium, Palm
  Coast, Fla., Feb. 25-Mar. 2, 1979, preprint CALT-68-709
  (retroprinted as hep-ph/9809459); S.~L.~Glashow, in “Proceedings of
  the Carg\'{e}se Summer Institute on Quarks and Leptons,”
  Carg\'{e}se, July 9-29, 1979, eds. M.~L\'{e}vy et. al, (Plenum,
  1980, New York) p707.  
  
\bibitem{Fukugita:1986hr} 
  M.~Fukugita and T.~Yanagida,
  Phys.\ Lett.\ B {\bf 174}, 45 (1986).
  doi:10.1016/0370-2693(86)91126-3

\bibitem{Nakayama:2016gvg} 
  K.~Nakayama, F.~Takahashi and T.~T.~Yanagida,
  Phys.\ Lett.\ B {\bf 757}, 32 (2016)
  doi:10.1016/j.physletb.2016.03.051
  [arXiv:1601.00192 [hep-ph]].
  
  
\bibitem{Esteban:2018azc} 
  NuFIT 4.0 (2018), www.nu-fit.org; I.~Esteban, M.~C.~Gonzalez-Garcia, A.~Hernandez-Cabezudo, M.~Maltoni and T.~Schwetz,
  JHEP {\bf 1901}, 106 (2019)
  doi:10.1007/JHEP01(2019)106
  [arXiv:1811.05487 [hep-ph]].

  
\bibitem{Bjorkeroth:2016qsk} 
  F.~Björkeroth, S.~F.~King, K.~Schmitz and T.~T.~Yanagida,
  Nucl.\ Phys.\ B {\bf 916}, 688 (2017)
  doi:10.1016/j.nuclphysb.2017.01.017
  [arXiv:1608.04911 [hep-ph]].

  
\bibitem{Casas:2001sr} 
  J.~A.~Casas and A.~Ibarra,
  Nucl.\ Phys.\ B {\bf 618}, 171 (2001)
  doi:10.1016/S0550-3213(01)00475-8
  [hep-ph/0103065].

\bibitem{Davidson:2002qv} 
  S.~Davidson and A.~Ibarra,
  Phys.\ Lett.\ B {\bf 535}, 25 (2002)
  doi:10.1016/S0370-2693(02)01735-5
  [hep-ph/0202239].


\bibitem{Mukaida:2012qn} 
  K.~Mukaida and K.~Nakayama,
  JCAP {\bf 1301}, 017 (2013)
  doi:10.1088/1475-7516/2013/01/017
  [arXiv:1208.3399 [hep-ph]].

\bibitem{Mukaida:2012bz} 
  K.~Mukaida and K.~Nakayama,
  JCAP {\bf 1303}, 002 (2013)
  doi:10.1088/1475-7516/2013/03/002
  [arXiv:1212.4985 [hep-ph]].
  
  
\bibitem{Buchmuller:2005eh} 
  W.~Buchmuller, R.~D.~Peccei and T.~Yanagida,
  Ann.\ Rev.\ Nucl.\ Part.\ Sci.\  {\bf 55}, 311 (2005)
  doi:10.1146/annurev.nucl.55.090704.151558
  [hep-ph/0502169].

\bibitem{Davidson:2008bu} 
  S.~Davidson, E.~Nardi and Y.~Nir,
  Phys.\ Rept.\  {\bf 466}, 105 (2008)
  doi:10.1016/j.physrep.2008.06.002
  [arXiv:0802.2962 [hep-ph]].

\bibitem{Giudice:2003jh} 
  G.~F.~Giudice, A.~Notari, M.~Raidal, A.~Riotto and A.~Strumia,
  Nucl.\ Phys.\ B {\bf 685}, 89 (2004)
  doi:10.1016/j.nuclphysb.2004.02.019
  [hep-ph/0310123].
  
\bibitem{Buchmuller:2004nz} 
  W.~Buchmuller, P.~Di Bari and M.~Plumacher,
  Annals Phys.\  {\bf 315}, 305 (2005)
  doi:10.1016/j.aop.2004.02.003
  [hep-ph/0401240].
  
 \bibitem{Plumacher:1996kc} 
  M.~Plumacher,
  Z.\ Phys.\ C {\bf 74}, 549 (1997)
  doi:10.1007/s002880050418
  [hep-ph/9604229].

\bibitem{KolbTurner}  
E.~W.~Kolb, M.~S.~Turner,
{\it The Early Universe}, Addison-Wesley, New York, 1990.

\bibitem{Plumacher:1997ru} 
  M.~Plumacher,
  Nucl.\ Phys.\ B {\bf 530}, 207 (1998)
  doi:10.1016/S0550-3213(98)00410-6
  [hep-ph/9704231].


\bibitem{Campbell:1992hd} 
  B.~A.~Campbell, S.~Davidson and K.~A.~Olive,
  Nucl.\ Phys.\ B {\bf 399}, 111 (1993)
  doi:10.1016/0550-3213(93)90619-Z
  [hep-ph/9302223].

\bibitem{Covi:1996wh} 
  L.~Covi, E.~Roulet and F.~Vissani,
  Phys.\ Lett.\ B {\bf 384}, 169 (1996)
  doi:10.1016/0370-2693(96)00817-9
  [hep-ph/9605319].

 \bibitem{Aghanim:2018eyx} 
  N.~Aghanim {\it et al.} [Planck Collaboration],
  arXiv:1807.06209 [astro-ph.CO].
  

\bibitem{Giudice:2004tc} 
  G.~F.~Giudice and A.~Romanino,
  Nucl.\ Phys.\ B {\bf 699}, 65 (2004)
  Erratum: [Nucl.\ Phys.\ B {\bf 706}, 487 (2005)]
  doi:10.1016/j.nuclphysb.2004.11.048, 10.1016/j.nuclphysb.2004.08.001
  [hep-ph/0406088].

\bibitem{Giudice:2011cg} 
  G.~F.~Giudice and A.~Strumia,
  Nucl.\ Phys.\ B {\bf 858}, 63 (2012)
  doi:10.1016/j.nuclphysb.2012.01.001
  [arXiv:1108.6077 [hep-ph]].

\bibitem{Murayama:1992ua} 
  H.~Murayama, H.~Suzuki, T.~Yanagida and J.~Yokoyama,
  Phys.\ Rev.\ Lett.\  {\bf 70}, 1912 (1993).
  doi:10.1103/PhysRevLett.70.1912
  
\bibitem{Murayama:1993xu} 
  H.~Murayama, H.~Suzuki, T.~Yanagida and J.~Yokoyama,
  Phys.\ Rev.\ D {\bf 50}, R2356 (1994)
  doi:10.1103/PhysRevD.50.R2356
  [hep-ph/9311326].
  
\bibitem{Murayama:1993em} 
  H.~Murayama and T.~Yanagida,
  Phys.\ Lett.\ B {\bf 322}, 349 (1994)
  doi:10.1016/0370-2693(94)91164-9
  [hep-ph/9310297].

\bibitem{Hamaguchi:2001gw} 
  K.~Hamaguchi, H.~Murayama and T.~Yanagida,
  Phys.\ Rev.\ D {\bf 65}, 043512 (2002)
  doi:10.1103/PhysRevD.65.043512
  [hep-ph/0109030].

\bibitem{Ellis:2003sq} 
  J.~R.~Ellis, M.~Raidal and T.~Yanagida,
  Phys.\ Lett.\ B {\bf 581}, 9 (2004)
  doi:10.1016/j.physletb.2003.11.029
  [hep-ph/0303242].
  
\bibitem{Antusch:2004hd} 
  S.~Antusch, M.~Bastero-Gil, S.~F.~King and Q.~Shafi,
  Phys.\ Rev.\ D {\bf 71}, 083519 (2005)
  doi:10.1103/PhysRevD.71.083519
  [hep-ph/0411298].

\bibitem{Antusch:2009ty} 
  S.~Antusch, M.~Bastero-Gil, K.~Dutta, S.~F.~King and P.~M.~Kostka,
  Phys.\ Lett.\ B {\bf 679}, 428 (2009)
  doi:10.1016/j.physletb.2009.08.022
  [arXiv:0905.0905 [hep-th]].

\bibitem{Kadota:2005mt} 
  K.~Kadota and J.~Yokoyama,
  Phys.\ Rev.\ D {\bf 73}, 043507 (2006)
  doi:10.1103/PhysRevD.73.043507
  [hep-ph/0512221].
  
\bibitem{Nakayama:2013nya} 
  K.~Nakayama, F.~Takahashi and T.~T.~Yanagida,
  Phys.\ Lett.\ B {\bf 730}, 24 (2014)
  doi:10.1016/j.physletb.2014.01.022
  [arXiv:1311.4253 [hep-ph]].

\bibitem{Binetruy:1987xj} 
  P.~Binetruy and M.~K.~Gaillard,
  Phys.\ Lett.\ B {\bf 195}, 382 (1987).
  doi:10.1016/0370-2693(87)90036-0

\bibitem{Gaillard:1995az} 
  M.~K.~Gaillard, H.~Murayama and K.~A.~Olive,
  Phys.\ Lett.\ B {\bf 355}, 71 (1995)
  doi:10.1016/0370-2693(95)00773-E
  [hep-ph/9504307].
  
\bibitem{Evans:2015mta} 
  J.~L.~Evans, T.~Gherghetta and M.~Peloso,
  Phys.\ Rev.\ D {\bf 92}, no. 2, 021303 (2015)
  doi:10.1103/PhysRevD.92.021303
  [arXiv:1501.06560 [hep-ph]].

\bibitem{Affleck:1984fy} 
  I.~Affleck and M.~Dine,
  Nucl.\ Phys.\ B {\bf 249}, 361 (1985).
  doi:10.1016/0550-3213(85)90021-5

\bibitem{Bolz:2000fu} 
  M.~Bolz, A.~Brandenburg and W.~Buchmuller,
  Nucl.\ Phys.\ B {\bf 606}, 518 (2001)
  Erratum: [Nucl.\ Phys.\ B {\bf 790}, 336 (2008)]
  doi:10.1016/S0550-3213(01)00132-8, 10.1016/j.nuclphysb.2007.09.020
  [hep-ph/0012052].

\bibitem{Pradler:2006qh} 
  J.~Pradler and F.~D.~Steffen,
  Phys.\ Rev.\ D {\bf 75}, 023509 (2007)
  doi:10.1103/PhysRevD.75.023509
  [hep-ph/0608344].

\bibitem{Pradler:2006hh} 
  J.~Pradler and F.~D.~Steffen,
  Phys.\ Lett.\ B {\bf 648}, 224 (2007)
  doi:10.1016/j.physletb.2007.02.072
  [hep-ph/0612291].

\bibitem{Rychkov:2007uq} 
  V.~S.~Rychkov and A.~Strumia,
  Phys.\ Rev.\ D {\bf 75}, 075011 (2007)
  doi:10.1103/PhysRevD.75.075011
  [hep-ph/0701104].

\bibitem{Asaka:2005cn} 
  T.~Asaka, K.~Ishiwata and T.~Moroi,
  Phys.\ Rev.\ D {\bf 73}, 051301 (2006)
  doi:10.1103/PhysRevD.73.051301
  [hep-ph/0512118].
  
\bibitem{Cheung:2011nn} 
  C.~Cheung, G.~Elor and L.~Hall,
  Phys.\ Rev.\ D {\bf 84}, 115021 (2011)
  doi:10.1103/PhysRevD.84.115021
  [arXiv:1103.4394 [hep-ph]].

\bibitem{Kawasaki:2006gs} 
  M.~Kawasaki, F.~Takahashi and T.~T.~Yanagida,
  Phys.\ Lett.\ B {\bf 638}, 8 (2006)
  doi:10.1016/j.physletb.2006.05.037
  [hep-ph/0603265].

\bibitem{Kawasaki:2006hm} 
  M.~Kawasaki, F.~Takahashi and T.~T.~Yanagida,
  Phys.\ Rev.\ D {\bf 74}, 043519 (2006)
  doi:10.1103/PhysRevD.74.043519
  [hep-ph/0605297].

\bibitem{Asaka:2006bv} 
  T.~Asaka, S.~Nakamura and M.~Yamaguchi,
  Phys.\ Rev.\ D {\bf 74}, 023520 (2006)
  doi:10.1103/PhysRevD.74.023520
  [hep-ph/0604132].
  
\bibitem{Endo:2006tf} 
  M.~Endo, K.~Hamaguchi and F.~Takahashi,
  Phys.\ Rev.\ D {\bf 74}, 023531 (2006)
  doi:10.1103/PhysRevD.74.023531
  [hep-ph/0605091].

\bibitem{Endo:2007ih} 
  M.~Endo, F.~Takahashi and T.~T.~Yanagida,
  Phys.\ Lett.\ B {\bf 658}, 236 (2008)
  doi:10.1016/j.physletb.2007.09.019
  [hep-ph/0701042].

\bibitem{Endo:2007sz} 
  M.~Endo, F.~Takahashi and T.~T.~Yanagida,
  Phys.\ Rev.\ D {\bf 76}, 083509 (2007)
  doi:10.1103/PhysRevD.76.083509
  [arXiv:0706.0986 [hep-ph]].
  
\bibitem{Hall:2012zp} 
  L.~J.~Hall, Y.~Nomura and S.~Shirai,
  JHEP {\bf 1301}, 036 (2013)
  doi:10.1007/JHEP01(2013)036
  [arXiv:1210.2395 [hep-ph]].
  
\bibitem{Hisano:2006nn} 
  J.~Hisano, S.~Matsumoto, M.~Nagai, O.~Saito and M.~Senami,
  Phys.\ Lett.\ B {\bf 646}, 34 (2007)
  doi:10.1016/j.physletb.2007.01.012
  [hep-ph/0610249].
  
\bibitem{Arvanitaki:2012ps} 
  A.~Arvanitaki, N.~Craig, S.~Dimopoulos and G.~Villadoro,
  JHEP {\bf 1302}, 126 (2013)
  doi:10.1007/JHEP02(2013)126
  [arXiv:1210.0555 [hep-ph]].

\bibitem{ArkaniHamed:2012gw} 
  N.~Arkani-Hamed, A.~Gupta, D.~E.~Kaplan, N.~Weiner and T.~Zorawski,
  arXiv:1212.6971 [hep-ph].

\bibitem{Ibe:2011aa} 
  M.~Ibe and T.~T.~Yanagida,
  Phys.\ Lett.\ B {\bf 709}, 374 (2012)
  doi:10.1016/j.physletb.2012.02.034
  [arXiv:1112.2462 [hep-ph]].

\bibitem{Ibe:2012hu} 
  M.~Ibe, S.~Matsumoto and T.~T.~Yanagida,
  Phys.\ Rev.\ D {\bf 85}, 095011 (2012)
  doi:10.1103/PhysRevD.85.095011
  [arXiv:1202.2253 [hep-ph]].

\bibitem{Hall:2011jd} 
  L.~J.~Hall and Y.~Nomura,
  JHEP {\bf 1201}, 082 (2012)
  doi:10.1007/JHEP01(2012)082
  [arXiv:1111.4519 [hep-ph]].

\bibitem{Ishiwata:2013waa} 
  K.~Ishiwata, K.~S.~Jeong and F.~Takahashi,
  JHEP {\bf 1402}, 062 (2014)
  doi:10.1007/JHEP02(2014)062
  [arXiv:1312.0954 [hep-ph]].

  
\bibitem{Hall:2009nd} 
  L.~J.~Hall and Y.~Nomura,
  JHEP {\bf 1003}, 076 (2010)
  doi:10.1007/JHEP03(2010)076
  [arXiv:0910.2235 [hep-ph]].

  
\bibitem{Hall:2014vga} 
  L.~J.~Hall, Y.~Nomura and S.~Shirai,
  JHEP {\bf 1406}, 137 (2014)
  doi:10.1007/JHEP06(2014)137
  [arXiv:1403.8138 [hep-ph]].

  
\end{thebibliography}
\end{document}